\documentclass[%
 reprint,
 amsmath,amssymb,
 aps,
]{revtex4-2}

\usepackage{graphicx}
\usepackage{dcolumn}
\usepackage{caption}  
\usepackage{ragged2e} 
\usepackage{bm}

\usepackage{algorithm}
\usepackage[noend]{algpseudocode}
\usepackage{placeins}
\usepackage{float} 

\usepackage{hyperref}
\hypersetup{
    colorlinks=true,
    linktoc=all,
    linkcolor=blue,
    citecolor=blue,
    urlcolor=blue
}

\usepackage{tikz}
\usepackage{subcaption}

\usetikzlibrary{shapes.geometric, arrows, positioning, calc}

\tikzstyle{startstop} = [rectangle, rounded corners, minimum width=4cm, minimum height=1.5cm, fill=black!10, text=black, font=\small, align=center]
\tikzstyle{process} = [rectangle, rounded corners, minimum width=4.5cm, minimum height=1.7cm, fill=blue!40, font=\small, align=center]
\tikzstyle{bottleneck} = [rectangle, rounded corners, minimum width=4.5cm, minimum height=1.7cm, fill=red!50, font=\small, align=center]
\tikzstyle{decoder} = [rectangle, rounded corners, minimum width=4.8cm, minimum height=1.7cm, fill=cyan!50, font=\small, align=center]
\tikzstyle{output} = [rectangle, rounded corners, minimum width=4cm, minimum height=1.5cm, fill=green!20, font=\small, align=center]
\tikzstyle{arrow} = [thick, ->, >=stealth, draw=black!80]
\tikzstyle{line} = [thick, draw=black!70]

\newcommand{\nh}[1]{{\color{black} {#1}}}

\begin{document}

\preprint{APS/123-QED}

\title{Deep source separation of overlapping gravitational-wave signals and non-stationary noise artifacts}%

\author{Niklas Houba}
\affiliation{ETH Zurich, Department of Physics, Institute for Particle and Astrophysics, Wolfgang-Pauli-Str.\ 27, 8093 Zurich, Switzerland}

\begin{abstract}
The Laser Interferometer Space Antenna (LISA) will observe gravitational waves in the millihertz frequency band, detecting signals from a vast number of astrophysical sources embedded in instrumental noise. Extracting individual signals from these overlapping contributions is a fundamental challenge in LISA data analysis and is traditionally addressed using computationally expensive stochastic Bayesian techniques. In this work, we present a deep learning-based framework for blind source separation in LISA data, employing an encoder-decoder architecture commonly used in digital audio processing to isolate individual signals within complex mixtures. Our approach enables signals from massive black-hole binaries, Galactic binaries, and instrumental glitches to be disentangled directly in a single step, circumventing the need for sequential source identification and subtraction. By learning clustered latent space representations, the framework provides a scalable alternative to conventional methods, with applications in both low-latency event detection and full-scale global-fit analyses. As a proof of concept, we assess the model's performance using simulated LISA data in a controlled setting with a limited number of overlapping sources. The results highlight deep source separation as a promising tool for LISA, paving the way for future extensions to more complex datasets.
\end{abstract}
\maketitle
\section{\label{sec:level1}Introduction}
The Laser Interferometer Space Antenna (LISA) is a space-borne gravitational-wave observatory developed by the European Space Agency (ESA) in collaboration with NASA, scheduled for launch in the mid-2030s \cite{colpi2024lisadefinitionstudyreport}. The detector consists of three spacecraft nominally arranged in an equilateral triangle, with each pair separated by 2.5 million kilometers \cite{Martens2021}. Using laser interferometry, LISA will measure fluctuations in spacetime caused by passing gravitational waves, extending the pioneering observations of ground-based detectors such as the Laser Interferometer Gravitational-Wave Observatory (LIGO) and Virgo \cite{PhysRevLett.116.061102, PhysRevLett.116.241103, PhysRevLett.118.221101, Abbott_2017, PhysRevLett.119.141101, PhysRevLett.119.161101, PhysRevX.9.031040, Abbott_2020, PhysRevD.102.043015, Abbott_2020_2, PhysRevLett.125.101102}. 

Unlike terrestrial detectors, which are constrained by seismic noise at low frequencies, LISA's space-borne configuration enables the detection of  gravitational waves in the 0.1 mHz to 1 Hz frequency band, a region densely populated with gravitational-wave sources \cite{SciReqDoc, Amaro-Seoane_2012}.

\subsection{Challenges in LISA data analysis}

LISA's sensitivity to millihertz sources will produce a data stream comprising a superposition of millions of overlapping gravitational-wave signals. Among these, Galactic binaries (GBs) -- and particularly double white-dwarf systems -- are expected to be so numerous that they will create an astrophysical noise floor, posing substantial challenges for scientific data analysis \cite{PhysRevD.75.043008, Criswell:2024hfn}. While a subset of sources, numbering in the tens of thousands, will be individually resolvable, the majority will blend into a persistent foreground noise, complicating the detection and characterization of other signals, including transient events from merging massive black-hole binaries (MBHBs). Instrumental noise further exacerbates the difficulty of disentangling individual sources \cite{Bayle2018b, PhysRevD.106.042005, Houba2023}. Addressing this challenge, known as the global-fit problem, requires high-performance, scalable data analysis algorithms capable of efficiently identifying and characterizing LISA's targets.

Solving the astrophysical global-fit problem requires methods that identify, model, and analyze gravitational-wave sources within a unified framework. Current approaches include Bayesian Markov Chain Monte Carlo (MCMC) and Maximum Likelihood Estimation (MLE) \cite{PhysRevD.111.024060, Littenberg2023PrototypeGA, PhysRevD.110.024005}. Both techniques have been successfully applied to simulated LISA datasets, each offering distinct trade-offs in computational cost, accuracy, and adaptability.  MCMC-based global fits, such as \texttt{Erebor} or \texttt{GLASS}, leverage ensemble sampling and GPU or parallel-CPU acceleration for improved efficiency \cite{PhysRevD.111.024060, Littenberg2023PrototypeGA}. These pipelines further rely on reversible-jump MCMC \cite{10.1093/biomet/82.4.711} to handle the uncertain number of sources in the data. To enhance computational efficiency, global-fit frameworks are structured to run large sampler modules covering a subset of sources in a blocked Gibbs fashion \cite{Ritter1992, purdue1991generic}. The approach ensures a consistent treatment of overlapping sources, but remains computationally demanding, potentially limiting its near-real-time application. 

\nh{In contrast, MLE-based methods typically follow a deterministic, step-wise signal extraction strategy, where sources such as MBHBs are estimated and subtracted before proceeding to fainter components like GBs \cite{PhysRevD.110.024005, Strub:2022upl}. While hierarchical subtraction is also employed in MCMC-based pipelines to enhance sampling efficiency, it is integrated within a broader Bayesian framework that jointly estimates all sources and parameters.}

Given these challenges, research into complementary approaches for source separation in LISA data remains an active and evolving field. Deep-learning  methods for data-driven feature extraction present a promising alternative by enabling direct source separation in a single step. These techniques offer advantages in computational efficiency, architectural flexibility, and scalability. Related work in the context of ground-based detectors, such as \texttt{DeepExtractor} \cite{dooney2025deepextractortimedomainreconstructionsignals}, has demonstrated the potential of deep learning for reconstructing gravitational-wave signals and mitigating transient noise artifacts. Moreover,  \texttt{UnMixFormer} \cite{Zhao_2025} has demonstrated the effectiveness of attention-based architectures for counting and separating overlapping compact binary coalescence signals in ground-based detector data.  Besides, simulation-based inference methods, such as Sequential Neural Likelihood \cite{efficientmassiveblackhole}, have recently been applied to LISA MBHB signals, enabling efficient posterior estimation with fewer simulator calls than traditional MCMC.

\nh{A key motivation for dedicated source separation and reconstruction stems from the fact that many gravitational-wave signals in LISA data overlap in both time and frequency, leading to strongly blended mixtures in the recorded data streams. This overlap poses a major obstacle for traditional Bayesian inference: the resulting likelihood surface becomes highly multimodal and degenerate, especially when multiple signals occupy the same frequency band. For example, accurately characterizing a faint GB becomes significantly more difficult when its signal is masked by a nearby, louder source -- whether of the same class or a different type. Without some form of source separation, classical parameter estimation methods must attempt to jointly fit overlapping signals, a process that is computationally expensive and scales poorly with source density.

Deep source separation addresses this problem by disentangling overlapping signals before parameter inference. This approach can transform the inference pipeline from a monolithic global fit into a modular two-stage process: (1) extract individual sources from the mixture and (2) perform parameter estimation on each extracted source independently or in smaller batches. As a result, source separation simplifies the inference landscape, reduces the dimensionality of the search space, and enables scalable parallelization.}

\subsection{Source separation in science and engineering}
 The task of untangling overlapping signals from a complex mixture remains both essential and challenging across various scientific and engineering disciplines \cite{Michelsanti2021, Vincent2018-li, Ochieng2023}. Imagine walking through a bustling city street, where car horns, music from storefronts, and conversations blend into a chaotic soundscape. While the human brain can effortlessly isolate specific voices or familiar sounds, digital audio processing struggles to achieve similar performance.

Early approaches leveraged statistical techniques such as Independent Component Analysis (ICA) to separate mixed signals mathematically \cite{Tharwat2020}. ICA operates under the assumption that the underlying sources are statistically independent, seeking a transformation that maximizes their separation. This is typically accomplished by expressing the observed mixed signals as a linear combination of unknown independent sources and estimating a separation matrix to recover the original signals without requiring prior knowledge of their specific characteristics. Beamforming methods, on the other hand, use microphone arrays to spatially isolate sound sources,  similar to how directional microphones enhance a speaker's voice in a noisy environment by focusing on sound from a specific direction while reducing background noise \cite{6661961}. More recently, deep learning has transformed source separation, enabling technologies such as music recognition systems that identify songs even in noisy environments \cite{Wang2003AnIS}, and AI-driven noise reduction in virtual meetings, which can intelligently distinguish speech from background interference in real-time \cite{lee2023speechenhancementvirtualmeetings}.

The city soundscape problem provides an intuitive analogy for source separation in LISA data. 
Just as city streets are filled with overlapping sounds that blend into a complex auditory scene, LISA's data stream is a cosmic symphony: gravitational waves from merging black holes and white-dwarf binaries overlap and mix with detector noise. Enter deep learning, which offers a data-driven approach to solving this astronomical puzzle. By utilizing robust architectures like convolutional and recurrent neural networks, deep-learning models can detect structured patterns hidden within the high-dimensional data, enabling scalable near-real-time blind source separation without prior knowledge of the number of sources \cite{ANSARI2023126895, Chien2018-no}. This paper marks the first step toward establishing deep source separation as a practical tool for LISA data \mbox{analysis}.

\subsection{Contribution and overview of the paper}
To address the challenge of source separation in LISA data, we introduce a deep learning-based framework designed to extract MBHBs, GBs, and instrumental glitches. Inspired by \texttt{demucs} (Deep extractor for music sources by Meta AI Research, see Ref.~\cite{defossez:hal-02379796}), a model originally developed for the separation of musical instruments in audio data, our approach replaces the traditional sequential subtraction paradigm in LISA data analysis with single-step source extraction.  \nh{While we do not reuse any code from \texttt{demucs}, we adopt a very similar architectural design}. By employing a shared encoder-decoder structure and latent space clustering, the model disentangles overlapping signals efficiently while remaining scalable to large source populations. A core feature of our framework is the frequency-binned output representation, which structures GB source separation on the basis of spectral content. This design helps mitigate source confusion by ensuring sources are dynamically clustered and disentangled in a learned feature space. Note that deep source separation operates independently of parameter estimation. This paper focuses on source separation, and parameter estimation based on its output is beyond the scope of this study.

The remainder of this paper is structured as follows. Section \ref{sec2} provides an overview of the expected LISA dataset and signal characteristics, including the types of gravitational-wave sources and the impact of time-delay interferometry (TDI) on data representation. Section \ref{sec3} describes our deep-learning framework, focusing on encoder-decoder architecture and latent space clustering. Section \ref{sec4} presents an evaluation of the model’s performance in extracting MBHBs, GBs, and glitches across various test scenarios. Finally, Section \ref{sec5} presents conclusions and proposes future work, including potential extensions to more complex astrophysical scenarios, and integration with full-scale LISA data pipelines.

\section{The LISA dataset}\label{sec2}
The dominant source populations in the millihertz LISA band are MBHBs and GBs, both of which present unique challenges for analyzing the LISA dataset. While MBHBs produce high-SNR transient signals, GBs form a persistent foreground that influences the detectability of other sources.

\vspace{-10pt}
\subsection{Massive black-hole binaries}
\begin{figure*}
    \centering
    \includegraphics[scale=0.85, clip, trim=35 510 45 80]{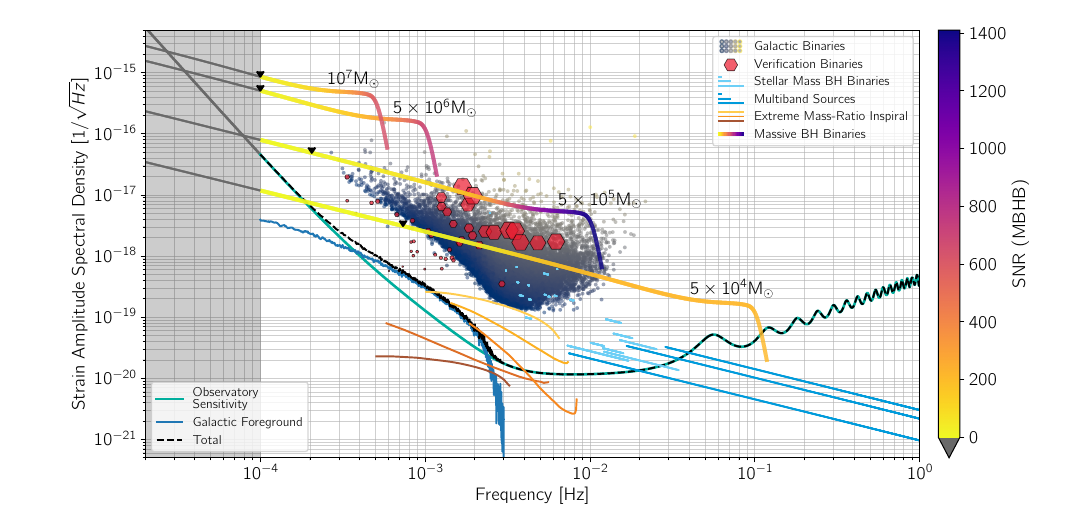}
    \caption{\RaggedRight Illustration of primary LISA source classes in the frequency-amplitude plane.  It includes merging massive black-hole binaries and extreme mass-ratio inspirals at moderate redshift, stellar-mass black holes at low redshift, and Galactic binaries, with sensitivity limits shown for instrumental noise, the unresolved Galactic foreground, and their sum. The cloud of resolvable sources appears above the noise level due to the detection threshold being set at an SNR of 7. Reprinted from \cite{colpi2024lisadefinitionstudyreport}.}
    \label{img:LISASources} 
\end{figure*}
MBHBs will be the loudest, most information-rich sources for LISA. They originate from the mergers of supermassive black holes at the centers of galaxies \cite{Mangiagli2022, mangiagli2024massiveblackholebinaries, Leroy2024}. These systems are expected to be detected across cosmic history, with events observable up to redshifts of $z \approx 15$. Their gravitational-wave emission sweeps through the LISA band as the binary inspirals toward coalescence, producing a high-SNR signal that lasts hours to weeks, depending on the total mass and redshift. MBHB detections will provide critical insights into black hole formation, galaxy evolution, and accretion physics.

As illustrated in Fig.~\ref{img:LISASources}, MBHBs with total masses between $10^4$ and $10^7$ $\mathrm{M}_\odot$ lie well within the LISA band, making them some of the loudest and most distant signals in LISA. \nh{Lower-mass massive black hole binaries (MBHBs) emit gravitational waves at higher frequencies and thus spend more time evolving within the LISA band before merger. In contrast, higher-mass systems have a lower merger frequency, often exiting the LISA sensitivity range before reaching its upper end, resulting in shorter in-band durations. Some MBHBs may also be multiband sources, entering the LISA band years before merger and later merging within the sensitivity window of ground-based detectors}

\vspace{-10pt}
\subsection{Galactic binaries}

Compact binaries in the Milky Way, particularly double white-dwarf systems, are expected to dominate the LISA band between 0.1 mHz and 10 mHz, producing nearly monochromatic individual signals that persist throughout the mission \cite{kupfer2024_lisa_gb, Cornish2017, Lackeos2023}. Unlike MBHBs, these binaries evolve slowly, with minimal frequency drift over LISA’s observational timescale. A subset of these binaries will be individually resolvable, particularly those with higher SNRs and well constrained parameters from electromagnetic observations. These verification binaries, depicted as red hexagons in Fig.~\ref{img:LISASources}, may be considered calibration sources for LISA, having been pre-identified through optical and radio surveys. However, this characterization remains under debate  \cite{littenberg2024lisaverificationbinariesfound}.
The vast majority of GBs will be unresolved, forming a stochastic foreground noise that dominates the low-frequency LISA band. This confusion-limited background, illustrated by the dashed black line in Fig.~\ref{img:LISASources}, limits LISA’s ability to detect fainter signals in the same frequency range, such as extreme mass-ratio inspirals (EMRIs) and a potential primordial gravitational-wave background. 

\nh{While this foreground is often modeled as stationary over short durations, it is in fact non-stationary on mission timescales. Two main mechanisms introduce this temporal evolution: (i) the intrinsic frequency drift of individual binaries due to gravitational radiation reaction, and (ii) the periodic Doppler modulation induced by LISA’s orbital motion around the Sun. These effects cause the apparent frequency and amplitude of sources to vary over time, imprinting slowly changing patterns on the composite foreground signal. On long timescales, such modulations can help distinguish overlapping sources by introducing characteristic time-frequency signatures that aid in source identification. Capturing these non-stationary features in data-driven models requires a large and diverse training sets that reflects the full range of time-dependent behavior expected during the mission.}

\subsection{Sources beyond the present study}

EMRIs are another important class of sources expected in the LISA band, resulting from the inspiral of a compact object -- typically a stellar-mass black hole, neutron star, or white dwarf -- into a much more massive black hole, usually found at the center of a galaxy \cite{PhysRevD.81.104014, PhysRevLett.133.141401, berry2019uniquepotentialextrememassratio}. These systems generate long-lived, complex waveforms as the smaller object undergoes tens of thousands of orbits before merging. EMRIs encode precise information about the spacetime geometry of the central massive black hole, making them key probes for testing strong-field General Relativity and the nature of black holes. EMRIs emit gravitational waves in the 1 mHz to 10 mHz range, overlapping with the Galactic foreground and some lower-mass MBHB signals. Their waveforms are highly intricate, containing multiple harmonics that encode information about the mass, spin, and orbital eccentricity of the system. Unlike MBHBs, which evolve rapidly through the LISA band, EMRIs remain in LISA’s sensitivity window for months to years. 
\nh{Accurately detecting, separating, and reconstructing such signals may require several methodological extensions to the framework presented in this paper, including, for example, hierarchical or multi-resolution network architectures, memory-aware encoders, or recurrent modules capable of capturing long-term temporal dependencies. Switching from raw time-domain inputs to time-frequency representations may also be beneficial, as they offer a more compact and structured view of slowly evolving signals like EMRIs. Additionally, due to their typically low signal-to-noise ratios, EMRIs are expected to require substantially larger training datasets to achieve reliable separation and reconstruction, which stands in contrast to the limited-data, proof-of-concept setting considered in this work. For these reasons, shared with current global-fit analyses that similarly omit EMRIs, we do not include them in the present study. The same applies to stellar-origin black hole binaries and unmodeled gravitational-wave bursts. Our current focus is on MBHBs, GBs, and non-stationary noise artifacts. Exploring the necessary architectural and data-driven adaptations remains an important direction for future research and will be essential to extending deep source separation methods to these challenging classes of sources.}

\vspace{-10pt}
\subsection{Instrumental noise and glitches}
In addition to astrophysical sources, the LISA data stream will contain instrumental noise and transient artifacts, both of which impact signal extraction and parameter estimation. These noise sources arise from multiple factors, including laser frequency fluctuations, unmodeled spacecraft acceleration, optical measurement noise, and environmental disturbances affecting the stability of the interferometric measurements \cite{Rdiger2008, PhysRevD.110.022003, BenderArt}.

One significant challenge is the presence of \textit{glitches}, short-duration noise transients caused by spacecraft systematics, or environmental perturbations, such as micrometeoroid impacts. These glitches can mimic or obscure real gravitational-wave signals, making their identification and mitigation essential for accurate source separation \cite{PhysRevD.105.042002, castelli2025extractiongravitationalwavesignals}. Characterizing instrumental noise is an active area of study, and techniques such as machine learning-based anomaly detection may play a crucial role in distinguishing true astrophysical signals from noise artifacts \cite{Houba2024}.

\vspace{-10pt}
\subsection{Time-delay interferometry}
Unlike ground-based detectors, which use simple Michelson interferometry, LISA's evolving geometry introduces unique challenges in maintaining phase coherence, requiring an advanced signal-processing technique known as TDI \cite{Tinto2002TDI1stGen, TDITINTO2005, PhysRevD.103.082001}. TDI is designed to suppress laser frequency noise, which would otherwise overwhelm gravitational-wave signals. Laser noise suppression is accomplished by linearly combining and time-shifting LISA's interferometric measurements to create virtual interferometers with equal arm lengths. It is important to note that the on-ground application of TDI transforms the data representation, altering the structure of signals compared to their raw strain measurements. In the context of machine learning, this requires that we train algorithms on TDI-processed data. Indeed, understanding these transformations is critical when designing traditional Bayesian inference or novel feature extraction methods.

\section{Framework for deep source separation of LISA signals}\label{sec3}
\nh{The source separation framework presented in this paper is designed to extract astrophysical signals and to identify and estimate glitches in the dataset.} By taking advantage of the model’s ability to learn structured latent representations, we can distinguish glitches from genuine gravitational wave events, reducing the risk of misclassification. This capability is particularly valuable in scenarios where glitches overlap with astrophysical signals, ensuring that transient artifacts do not interfere with the accurate reconstruction of MBHBs or GBs.

The following section provides an overview of key deep learning-based methods for source separation, highlighting their strengths and applications. This is followed by a detailed presentation of the framework developed in this work for LISA data analysis.

\subsection{Overview of deep-learning techniques for source separation}

A widely adopted approach in source separation involves \textit{mask-based methods}, where neural networks are trained to estimate time-frequency masks that enhance the separation of individual sources when applied to the spectrogram of an input mixture \cite{li2022usedeepmaskestimation, Simpson2015ProbabilisticBC}. Typically, such architectures consist of a neural network that processes the magnitude spectrogram through layers of batch normalization, multiple bi-directional long short-term memory (BLSTM) networks, and a fully connected output layer with a sigmoid activation function to generate the masks. The network is trained using a reconstruction loss, commonly an L1 or L2 loss between the estimated and target spectrograms. Variations of mask-based methods include soft masking, where estimated masks take continuous values between 0 and 1, and hard masking, where values are binarized. This approach is particularly effective in speech separation and enhancement, as it leverages the structured nature of human speech signals.

\textit{Deep clustering} presents an alternative approach, addressing source separation as an embedding-based learning problem \cite{10.1007/978-3-030-70665-4_117, Li2024, 10.5555/3045390.3045442}. Instead of estimating masks directly, deep clustering models learn to map each time-frequency bin of the input spectrogram into a high-dimensional embedding space. In this space, embeddings corresponding to the same source cluster together, while those from different sources remain well-separated. Clustering algorithms, such as $k$-means \cite{MORADIFARD2020185}, are subsequently applied to assign time-frequency bins to their respective sources and generate separation masks. This method has shown superior performance in tasks such as blind source separation and reverberant speech separation, where the relationship between sources is highly nonlinear.

\textit{Chimera networks} are hybrid architectures that integrate both mask-based and deep clustering techniques within a unified framework through multi-task learning \cite{Luo2016DeepCA, Chang2019, 9054340}. These networks contain shared BLSTM layers, followed by dual output heads: one for deep clustering and another for mask inference. During training, the deep clustering objective serves as a regularizer, enhancing the generalization capability of the network, while the mask inference objective is used for direct source separation during inference. Chimera networks have demonstrated improved robustness in real-world conditions, benefiting from the complementary strengths of both deep clustering and mask-based learning.

While source separation traditionally operates on spectrogram representations, time-domain approaches have emerged as a powerful alternative, enabling direct processing of raw audio waveforms \cite{9403999, 9053934}. Time-domain models have demonstrated state-of-the-art performance, surpassing traditional spectrogram-based approaches in many benchmarks due to their ability to preserve phase information and reduce artifacts introduced by spectral transformations. Notable architectures in this category include Conv-TasNet \cite{10.1109/TASLP.2019.2915167}, a convolutional time-domain audio separation network that employs an encoder-decoder structure with temporal convolutional networks. By replacing the conventional short-time Fourier transform with a learned encoder, Conv-TasNet captures fine-grained temporal structures, enhancing speech separation quality. Another prominent model, \texttt{demucs} \cite{defossez:hal-02379796}, is inspired by deep generative models for audio and features a U-Net-like architecture \cite{UNet} with a convolutional encoder, a BLSTM-based bottleneck and a decoder utilizing transposed convolutions. This design effectively captures both local and long-range temporal dependencies, making \texttt{demucs} particularly well-suited for music source separation tasks, where harmonic and percussive elements are intertwined. In this work, we employ a modified \texttt{demucs}-based encoder-decoder network. We outline its mathematical theory in the next section.

\subsection{Encoder-decoder architectures}

In this paper, deep learning-based source separation employs an encoder-decoder architecture, encoding raw input signals into a compressed latent representation before reconstructing the individual components. Unlike spectrogram-based methods, time-domain approaches naturally {preserve phase information} and 	 {reduce spectral artifacts}, which is important for signal reconstruction in high-dimensional gravitational wave data. In the context of LISA, the encoder processes a noisy mixture and extracts the most important patterns and features, transforming the raw input into a more structured form. At the core of this process is the \textit{bottleneck}, a stage where information is temporarily compressed, ensuring that only the most relevant details are retained while filtering out noise and redundancies. The bottleneck representation helps the model focus on essential aspects of the data, improving the separation of different sources. Finally, the decoder uses this refined information to reconstruct the individual signals corresponding to MBHBs, GBs, and glitches. This section introduces the mathematical foundations behind this framework and explains how it helps disentangle overlapping signals effectively.

\subsubsection{Encoder}

The shared encoder is responsible for mapping raw LISA TDI data into a structured latent space that highlights key features relevant to source separation. The term ``shared'' refers to the fact that a single encoder processes the entire input mixture and extracts a common feature representation, which is then used by multiple decoders to reconstruct individual sources. Instead of training separate encoders for each source type, a shared encoder ensures unified feature extraction, improving efficiency and consistency in learned representations.

Given a raw time-domain signal $x(t)$, the encoder function can be formulated as

\begin{equation}
    z = E(x; \theta_E),
\end{equation}
where $z$ represents the latent space encoding that captures essential waveform structures, while $E(\cdot; \theta_E)$ denotes the encoder network parameterized by $\theta_E$. The encoder typically comprises multiple convolutional layers to extract local time-frequency patterns, followed by nonlinear activations to improve source separability. The parameters $\theta_E$ are learned through training.

\subsubsection{Latent representation and bottleneck transformation}

The latent space provides a compact representation of extracted features, facilitating the separation of individual sources. In the context of blind source separation, it enables the mapping of overlapping signals to distinct regions, aiding in their disentanglement and improving reconstruction accuracy. The transformation reduces redundancy, ensuring that the model focuses on independent components.
To further refine the extracted features and prevent the network from overfitting, an additional constraint is introduced through the bottleneck layer. The bottleneck layer serves as a regularization mechanism, limiting the amount of information passing through the network. It ensures that only the most relevant features are retained while suppressing noise and redundant details. This process can be expressed as
\begin{equation}
    \tilde{z} = B(z; \theta_B),
\end{equation}
where $\tilde{z}$ represents the bottleneck encoding, $g(\,\cdot\,; \theta_B)$ is the low-dimensional projection that filters irrelevant components while preserving key signal characteristics required for reconstruction and $\theta_B$ represents the trainable parameters of the bottleneck function.

A more rigorous way to understand the bottleneck transformation is through information theory, where the goal is to find a representation $\tilde{z}$ that retains as much relevant information about the original signal $x(t)$ as possible while discarding unnecessary details (e.g., noise and redundant components). This is captured by the \textit{information bottleneck objective} \cite{tishby2000informationbottleneckmethod}, which aims to optimize the trade-off between compression and preservation of useful information:
\begin{equation}
    \max_{\theta_B} \quad I(\tilde{z}; x) - \beta I(\tilde{z}; n).\label{eq:BottObj}
\end{equation}
\nh{Here, $I(A; B)$ denotes the mutual information between variables $A$ and $B$, quantifying how much knowing one reduces uncertainty about the other. The term $I(\tilde{z}; x)$ ensures that the compressed representation retains meaningful information about the input, while $I(\tilde{z}; n)$ penalizes the retention of irrelevant information. In our context, $n$ corresponds to components of the input that are not meant to be explicitly reconstructed, primarily the quasi-stationary instrumental noise. To guide the encoder and bottleneck toward discarding irrelevant components, we use a frequency-dependent noise model to generate diverse time-domain noise realizations during training. This exposure enables the network to distinguish between signals of interest and stationary noise, and to focus its representational capacity on features relevant to signal reconstruction.

It remains an open question how performance is affected when the evaluation data exhibits noise properties that differ from the training distribution. Note that this is not a specific limitation of our method, but a general challenge in machine learning-based analyses of noisy, high-dimensional measurements.}

\subsubsection{Decoder}

The decoder reconstructs individual sources from the shared latent space by applying learned transformations. Each decoder head receives the same latent input but is trained to reconstruct only a specific target source:

\begin{equation}
    \hat{x}_i = D_i(\tilde{z}; \theta_{D_i}),
\end{equation}
where $\hat{x}_i$ is the reconstructed output for the $i$-th source, and $D_i(\,\cdot\,; \theta_{D_i})$ represents the decoder network parameterized by $\theta_{D_i}$, responsible for reconstructing individual components. The decoder can apply a series of transposed convolutions to progressively upsample and restore temporal structures from the compressed latent representation. To enhance reconstruction accuracy, \textit{skip connections} can be incorporated, allowing the network to retain fine-grained details by reintroducing relevant features from earlier encoding layers. Typically, source-specific activation functions are employed to ensure that each decoder head reconstructs only its assigned target, preventing interference between different signal types.

\subsection{\texttt{Demucs} as an example of an established encoder-decoder model}
After introducing the fundamental principles of encoder-decoder architectures and latent space representations, we now turn our attention to \texttt{demucs}. The model has demonstrated success in audio source separation tasks and will be adapted for gravitational wave data analysis in the context of LISA.

Unlike spectrogram-based approaches that rely on time-frequency representations, \texttt{demucs} operates directly on raw audio waveforms, allowing the model to fully leverage the temporal and structural characteristics of sound, resulting in improved separation performance \cite{defossez:hal-02379796}. At its core, \texttt{demucs} is based on a U-Net convolutional architecture. U-Net consists of an encoder-decoder framework with symmetric skip connections that link corresponding layers between the encoder and decoder paths \cite{10.1007/978-3-319-24574-4_28}. Note that skip connections mitigate the bottleneck's impact by reintroducing high-resolution features. In \texttt{demucs}, where maintaining the temporal structure of waveforms is essential, these connections help preserve fine details that might otherwise be lost during compression.

The encoder in \texttt{demucs} comprises multiple convolutional layers that progressively downsample the input, capturing hierarchical features. Each convolutional block integrates standard convolutions, batch normalization, and nonlinear activation functions to enhance feature extraction. \nh{To improve its capacity to model temporal dependencies within each input segment, \texttt{demucs} incorporates BLSTM layers within the bottleneck. These allow the model to process both forward and backward temporal context over the segment duration. This is particularly advantageous for disentangling overlapping musical components that exhibit strong temporal structure -- such as harmonically related instruments or time-aligned effects -- within the scope of each training snippet.} Additionally, \texttt{demucs} employs gated linear units as activation functions, which enhance the model’s expressiveness by selectively regulating information flow.

The decoder path in \texttt{demucs} employs transposed convolutional layers to upsample the encoded features, reconstructing the separated sources while preserving their fine-grained temporal structure. The inclusion of skip connections from the encoder to the decoder ensures that high-resolution details lost during downsampling are retained, leading to accurate reconstruction of the separated signals. 

\texttt{Demucs} integrates several additional techniques to improve performance. Its multi-scale processing capability, facilitated by the hierarchical convolutional structure, enables the model to capture both short-term transients and long-term harmonic structures. Although \texttt{demucs} is trained on short audio snippets, it can process long audio sequences effectively by incorporating \textit{overlapping window inference}. This approach involves applying a sliding window with overlapping segments, which not only mitigates boundary artifacts but also ensures smooth transitions between separated chunks, thereby preserving temporal consistency over extended durations. 

\subsection{Modifying \texttt{demucs} for application in LISA}
 
Adapting \texttt{demucs} for LISA data requires modifications that account for the unique characteristics of gravitational-wave signals. Unlike conventional audio streams, LISA data comprises a superposition of overlapping astrophysical waveforms and transient instrumental glitches. The primary challenge in this adaptation lies in accurately isolating individual sources while ensuring the scalability of the separation model. To address this, we design a modified architecture to structure the separation task across three dedicated decoder heads following a shared encoder and bottleneck representation. The first decoder is responsible for reconstructing MBHB signals, assuming for simplicity that individual MBHB events do not overlap in time. The second decoder isolates non-stationary noise artifacts, such as transient glitches. The third decoder is specifically designed for GBs and consists of multiple output channels, each corresponding to a predefined frequency bin. This multi-output design enables the model to disentangle individual GB sources while maintaining scalability. Instead of assigning a separate decoder to each GB source -- an approach that quickly becomes computationally prohibitive -- the frequency space is discretized into small bins, ensuring that each bin contains at most one dominant source. \nh{This mirrors the assumption used for MBHBs in the time domain, where each time chunk is assumed to contain at most one MBHB signal.}
In our framework, we intentionally omit skip connections to maintain simplicity. As a result, our bottleneck applies a weaker compression compared to \texttt{demucs}. Details will follow.

\nh{We use LISA's TDI data streams A, E, and T \cite{Prince2002} as input to the model. These channels form an orthogonal and widely used basis that spans the full gravitational-wave response space of the detector; any other complete set of TDI combinations would be equally valid from an information perspective. In the current setup, each TDI channel is processed independently using a shared separation model. This simplification reduces model complexity and training cost, but it discards potentially useful cross-channel correlations. As a result, source reconstructions across the channels may become inconsistent in the presence of non-stationary noise or low signal-to-noise ratios. Future extensions of the framework will adopt joint multi-channel processing -- such as through shared or cross-channel encoder structures -- to better leverage the complementarity of different TDI observables. While such modifications may improve reconstruction accuracy and consistency, they are not essential for the proof-of-concept separation task presented here. Ultimately, training the network to recover the underlying gravitational-wave strain in the barycentric frame, e.g., the \( h_+ \) and \( h_\times \) polarizations, could offer further advantages for downstream parameter estimation and is left for future work.} The schematic architecture of the \texttt{demucs}-inspired multi-source extraction model is depicted in Fig.~\ref{fig:ModDemucsLISA}.
\begin{figure*}[t!]
    \centering
\begin{center}
\scalebox{0.75}{ 
\begin{tikzpicture}[node distance=1.5cm] 

    \node (input) [startstop] {Noisy LISA data $x(t)$ \\ \small (TDI space)};

    \node (encoder) [process, below=of input] {Shared encoder 
    \\ \(\displaystyle z = E(x; \theta_E)\)};
    
    \node (bottleneck) [bottleneck, below=of encoder] {Bottleneck \\ \(\displaystyle \tilde{z} = B(z; \theta_B)\)};

    \node (mbhb_decoder) [decoder, below=of bottleneck, xshift=-5.5cm, yshift=-0.2cm] {MBHB decoder  \\  \(\displaystyle \hat{x}_{\mathrm{MBHB}} = D_{\mathrm{MBHB}}(\tilde{z}; \theta_{D_{\mathrm{MBHB}}})\)};
    
    \node (glitch_decoder) [decoder, below=of bottleneck, xshift=0cm, yshift=-0.2cm] {Glitch decoder  \\ \(\displaystyle \hat{x}_{\mathrm{glitch}} = D_{\mathrm{glitch}}(\tilde{z}; \theta_{D_{\mathrm{glitch}}})\)};

    \node (gb_decoder) [decoder, below=of bottleneck, xshift=5.5cm, yshift=-0.2cm] {GB decoder  \\ \small \(\displaystyle \hat{x}_{\mathrm{GB},i} = D_{\mathrm{GB}}(\tilde{z}; \theta_{D_{\mathrm{GB}}})\)\\ \small (multi-channel output)};

    \node (mbhb_out) [startstop, below=of mbhb_decoder, yshift=-0.1cm] {Extracted MBHB signal \\ \small (TDI space)};
    \node (glitch_out) [startstop, below=of glitch_decoder, yshift=-0.1cm] {Extracted glitch signal \\ \small (TDI space)};
    \node (gb_out) [startstop, below=of gb_decoder, yshift=-0.1cm] {Extracted GB signals \\ \small (TDI space)};

    \draw [arrow] (input) -- (encoder) node[midway, right] {Feature extraction};
    \draw [arrow] (encoder) -- (bottleneck) node[midway, right] {Feature compression};

    \draw [line] (bottleneck) --  ++(0,-2.0) coordinate (split) node[midway, right] {Latent representation};

    \draw [arrow] (split) -| (mbhb_decoder.north);
    \draw [arrow] (split) -- (glitch_decoder.north);
    \draw [arrow] (split) -| (gb_decoder.north);

    \draw [arrow] (mbhb_decoder.south) -- (mbhb_out.north) node[midway, right] {Reconstruction};
    \draw [arrow] (glitch_decoder.south) -- (glitch_out.north) node[midway, right] {Reconstruction};
    \draw [arrow] (gb_decoder.south) -- (gb_out.north) node[midway, right] {Reconstruction};
\end{tikzpicture}
}
\end{center}
    \caption{\RaggedRight Deep source separation framework for LISA data, where a shared encoder compresses the TDI input, and decoders reconstruct MBHBs, GBs and glitches. Since the input data denotes a TDI channel, the separated and decoded output signals are represented in the TDI space, as well.}
    \label{fig:ModDemucsLISA}
\end{figure*}

\subsubsection{Encoder and bottleneck architecture of LISA-modified \texttt{demucs}}

The encoding process transforms the noisy TDI time series containing overlapping MBHBs, GBs, and glitches into a structured latent representation. The encoder consists of four consecutive one-dimensional convolutional layers with increasing feature dimensions, each followed by a ReLU activation to introduce nonlinearity. This hierarchical feature extraction progressively captures waveform structures at different resolutions, preserving temporal patterns. Once the latent representation $z$ is obtained, a bottleneck layer refines the extracted features by applying an additional one-dimensional convolution. Here, the bottleneck transformation restructures the learned representations by reducing the number of feature channels rather than compressing the sequence length. The encoder follows a shared architecture, utilizing a single feature extraction pipeline for all sources. Table \ref{tab:encoder_spec} summarizes the network configuration.

\begin{table}
    \centering
    \caption{\RaggedRight Encoder and bottleneck network configuration. The padding parameters are chosen to ensure that, when combined with the decoder, the framework's outputs match the dimension of the input signal.}
    \begin{tabular}{c c c c c c}
        \hline
        Layer & Input & Output & Kernel & Stride & Padding \\
        \hline
        Conv1D & 1 & 64 & 8 & 2 & 3 \\
        ReLU   & - & - & - & - & - \\[2pt]
        Conv1D & 64 & 128 & 8 & 2 & 3 \\
        ReLU   & - & - & - & - & - \\[2pt]
        Conv1D & 128 & 256 & 8 & 2 & 3 \\
        ReLU   & - & - & - & - & - \\
        Conv1D & 256 & 512 & 9 & 1 & 4 \\
        ReLU   & - & - & - & - & - \\[4pt]
        Bottleneck & 512 & 256 & 3 & 1 & 1 \\
        \hline
    \end{tabular}%
    \label{tab:encoder_spec}
\end{table}

Since the first three convolutional layers in the encoder apply a stride of 2 each, the input time series is progressively downsampled by a factor of \(2^3 = 8\), meaning the sequence length is reduced to \(1/8\) of the original input. Note that the bottleneck layer does not further compress the sequence length because it applies a convolution with stride 1. Instead, it reduces the number of feature channels from 512 to 256, serving as a feature refinement step rather than a strict compression bottleneck. Unlike \texttt{demucs}, which uses skip connections to preserve fine-grained details, our encoder does not include skip connections for now. Therefore, we decided that the bottleneck layer should preserve the temporal resolution while focusing on channel-wise dimensionality reduction. In future work, we plan to experiment with the combination of skip connections and bottleneck compression to assess their impact on feature retention and downstream tasks.

In Section \ref{sec4}, we will visualize the bottleneck-encoded latent features using $t$-distributed stochastic neighbor embedding ($t$-SNE) \cite{JMLR:v23:21-0524}, a dimensionality reduction technique that maps high-dimensional data into a lower-dimensional space to illustrate the clustering patterns in typical LISA TDI data.

\vspace{-5pt}
\subsubsection{Multi-source decoder heads for MBHBs, GBs, and glitches}

The decoder reconstructs individual gravitational-wave signals from the shared latent representation by applying a dedicated decoding process for each source type. It follows a transposed convolutional architecture similar to \texttt{demucs}, where the extracted latent features are progressively upsampled back into de-noised time-domain waveforms. Each decoder consists of three transposed convolution layers with ReLU activations, allowing for structured reconstruction of the signals.

For MBHB and glitch signals, the decoder reconstructs the time-domain waveform as a single-channel output. This design choice assumes for simplicity that MBHB merger signals occur at sufficiently distinct times, eliminating the need for explicit separation. Additionally, instead of resolving individual glitches, the model treats them as a collective class and outputs a single data stream that may contain multiple glitches.

For GB sources, we adopt a different approach. The decoder utilizes a frequency-binned method, generating a multi-channel waveform where each channel corresponds to a distinct frequency bin. Each bin reconstructs the portion of the GB compound that falls within its designated frequency range. This design eliminates source permutation ambiguity and allows the model to accommodate an arbitrary number of GB sources efficiently. The implementation features a final transposed convolution layer with $N_{\text{GB}}$  output channels, where $N_{\text{GB}}$  denotes the predefined number of frequency bins. The decoder processes the latent representation in a single forward pass, simultaneously producing a time-domain waveform for each bin. If a bin contains no GB sources, the model naturally learns to suppress that output channel. 

Table \ref{tab:decoder_spec} summarizes the architecture of the decoder networks used for reconstructing MBHB, GB, and glitch waveforms. Note that the padding parameters are selected to ensure that the decoded signals retain the same length as the input signal prior to encoding.

\begin{table}
    \centering
    \caption{\RaggedRight Decoder network configuration. The output channel dimension $c$ is $1$ for MBHBs and glitches, while it is set to $N_{\text{GB}}$ for GBs, where each channel corresponds to a frequency bin. Note that ConvTranspose1D performs the inverse operation of Conv1D.}
    \begin{tabular}{c c c c c c}
        \hline
        Layer & Input & Output & Kernel & Stride & Padding \\
        \hline
        ConvTranspose1D & 256 & 128 & 8 & 2 & 3 \\
        ReLU   & - & - & - & - & - \\[2pt]
        ConvTranspose1D & 128 & 64 & 8 & 2 & 3 \\
        ReLU   & - & - & - & - & - \\[2pt]
        ConvTranspose1D & 64 & $c$ & 8 & 2 & 4+1\footnote{An additional output padding of 1 is applied to ensure the decoded signals match the input signal length, which is required for computing the loss function.} \\
        \hline
    \end{tabular}%
    \label{tab:decoder_spec}
\end{table}

Each decoder (MBHB, glitch, and GB) follows this same transposed convolutional architecture, upsampling the latent representation back to the time-domain waveform. The structure ensures that the output sequence length matches the original input length after three layers of transposed convolution with stride 2. This preserves the temporal coherence of the reconstructed signals. By using a frequency-binned approach for GB separation, the model effectively assigns sources to their nearest bin based on frequency content. This strategy eliminates the need for a predefined number of GB sources and ensures scalability to thousands of sources.

In the simulation section of this paper, we will use a low value for $N_{\text{GB}}$ and explore more realistic populations in a future study by increasing the number of hidden layers, neurons, and training data size.
\vspace{-13pt}

\subsubsection{Loss calculation and training}

The model is trained by optimizing a total loss function that consists of individual Mean Squared Error (MSE) loss terms for MBHBs, GBs, and glitches. The loss is computed based on the reconstructed waveforms produced by the decoder and their corresponding ground truth signals. The total loss function is defined as
\begin{equation}
    L_{\text{Total}} = L_{\text{MBHB}} +  L_{\text{glitch}} + L_{\text{GB}}, \label{eq:TotalLoss}
\end{equation}
where $L_{\text{MBHB}}$, $L_{\text{glitch}}$, and $L_{\text{GB}}$ represent the reconstruction losses for each source type.

For MBHB and glitch signals, the loss is computed directly by comparing the predicted output $\hat{x}$ with the ground truth ${x}$ using MSE:
\begin{equation}
    L_{\text{MBHB}} = \text{MSELoss}(\hat{x}_{\text{MBHB}}, x_{\text{MBHB}})
\end{equation}
and
\begin{equation}
    L_{\text{glitch}} = \text{MSELoss}(\hat{x}_{\text{glitch}}, x_{\text{glitch}}).
\end{equation}
\nh{For GB sources, a frequency-binned representation is used to ensure structured separation of the overlapping signals in the spectral domain. Each GB signal is assigned to its nearest frequency bin based on its central frequency. Formally, for each batch $b$ and bin index $k$, the binned GB signal is computed as
\begin{equation}
    x_{\text{GB}}^{\text{bin}}[b, k, t] = \sum_{i \in \mathcal{I}_{b, k}} x_{\text{GB}}[b, i, t],
    \label{eq:gb_bin_assign}
\end{equation}
where $x_{\text{GB}}[b, i, t]$ denotes the waveform of the $i$-th GB source and
\begin{equation}
    \mathcal{I}_{b, k} = \left\{ i \,\middle|\, \arg\min_j \left| f_{\text{GB}}[b, i] - f_{\text{bins}}[j] \right| = k \right\}
\end{equation}
is the set of sources assigned to bin $k$ based on their frequency proximity to the predefined bin centers $f_{\text{bins}}$.} The frequency-binned ground
truth signal is then compared to the predicted GB output using MSE:
\begin{equation}
    L_{\text{GB}} = \text{MSELoss}(\hat{x}_{\text{GB}}, x_{\text{GB}}^{\text{bin}}).
\end{equation}
For this strategy to resolve each GB individually, the bin width must be small enough to ensure that each bin contains at most one GB source. If the bins are too large, multiple sources will be mapped to the same bin, leading to signal blending. Conversely, if there are fewer GB sources than bins, the model suppresses the corresponding outputs, maintaining efficiency while ensuring scalability to thousands of sources without requiring prior knowledge of the actual number of GB signals in the data.

After computing the total loss, gradients are propagated backward through the network, and model parameters are updated using the Adam optimizer. The optimization process follows standard backpropagation, iteratively refining the trainable parameters of the encoder, bottleneck, and decoder networks. The training loop follows these key steps:

\begin{enumerate}
    \item The model receives mixed gravitational-wave signals as input and predicts the MBHB, glitch, and GB outputs.
    \item The loss is computed separately for MBHBs, glitches, and frequency-binned GBs.
    \item The gradients are computed using backpropagation, and model parameters are updated using the Adam optimizer.
    \item The process is repeated for multiple training epochs, progressively improving the model’s ability to disentangle overlapping signals.
\end{enumerate}
\nh{Given that LISA observations span months to years, the model must be capable of processing long-duration signals while preserving temporal coherence. However, training directly on full-length time-series data is computationally prohibitive due to the memory and processing demands of handling inputs with millions of time steps. Such long sequences would exceed the memory limits of standard hardware, particularly when used in conjunction with deep convolutional encoder-decoder architectures.

To address this, we adopt an approach similar to that used in \texttt{demucs}, where the model is trained on randomly sampled short-duration segments (e.g., minutes to hours). This strategy significantly reduces memory usage, allows for efficient batch processing, and promotes generalization. When applying the model to full-length LISA data, it will be necessary to divide the time series into overlapping segments, process each independently, and then merge the outputs using a weighted averaging scheme. This stitching mechanism -- also employed in \texttt{demucs} -- ensures continuity across segment boundaries and mitigates edge artifacts.} This work will be presented in a follow-up paper.

Note that the original \texttt{demucs} architecture is more complex than the implementation used in this study. It features a deeper network structure with additional convolutional layers, a larger number of feature channels, and BLSTM layers to model long-range dependencies in audio waveforms. In contrast, our implementation adopts a simplified architecture with fewer layers and reduced model complexity, focusing primarily on demonstrating the feasibility of deep blind source separation in LISA's TDI data. While we currently maintain this streamlined model for proof-of-concept simulations, it is possible to further converge toward the full \texttt{demucs} architecture by increasing network depth, adding additional hidden layers, or expanding the number of neurons in each processing stage. The flexibility of the chosen framework ensures that extensions are feasible, enabling the method to be progressively adapted for larger and more intricate LISA data analysis tasks. As the scope of this study is not to resolve the full source population expected in LISA but rather to demonstrate the viability of deep learning-based source separation in a controlled setting with a limited number of sources, we stick to the architectural design proposed in this section when presenting the simulation results in the following. The simplified design already proves effective in achieving remarkable separation of individual signals, suggesting that deep learning-based source separation can offer a more streamlined approach to future pipeline implementation.

\nh{
\subsection{Future directions for resolving overlapping Galactic binaries}

The current GB decoder prototype is based on a simplifying assumption: that the frequency axis can be discretized into narrow bins such that each bin contains at most one GB signal. In practice, the spectral density of the GB population will be high, especially at low frequencies, making this assumption increasingly fragile.
In densely populated regions of the spectrum, multiple GB sources with closely spaced frequencies may fall within the same bin, even under carefully optimized decoder binning schemes. These near-degenerate signals, though individually narrow-band, can interfere constructively or destructively, particularly when their amplitudes are similar, posing a fundamental challenge for source separation in LISA. While differences in parameters such as sky position can, in principle, induce distinct Doppler and amplitude modulation patterns that aid disentanglement, the current purely bin-wise decoder design lacks the capacity to fully exploit such subtle variations. As a result, the model may struggle to accurately resolve individual signals in frequency-overlapping scenarios -- even if their SNRs would allow distinguishability in theory.

A full treatment of this issue lies beyond the scope of the present study. Nevertheless, we outline some architectural directions that move beyond the current one-source-per-bin assumption. One promising strategy is to adopt a dynamic multi-slot decoding scheme, in which each frequency bin produces a flexible number of output slots, with each slot representing a distinct candidate source. Rather than fixing the number of slots a priori, mechanisms such as \emph{Slot Attention} \cite{10.5555/3495724.3496691} or set-based transformers could iteratively infer both the number and identity of sources present, conditioned on local and global features. This approach would allow the model to adaptively allocate representational capacity based on local source density. To better resolve sources whose features span multiple bins (e.g., due to modulation or frequency drift), future architectures could also incorporate sequence modeling across bins, using temporal convolution, recurrent layers, or attention-based modules. These models can exploit correlations between neighboring bins to disentangle overlapping signals that cannot be separated using purely local information. Training such models would require a combination of reconstruction loss, permutation-invariant supervision (e.g., using the Hungarian algorithm \cite{Kuhn1955}), and disentanglement-promoting regularizers (such as orthogonality penalties or contrastive objectives). 

Future work will evaluate the effectiveness of this strategy on realistically dense GB populations, where overlapping sources are not rare exceptions but rather a fundamental aspect of the data.

}

\section{Simulation results}\label{sec4}

The section presents the results of the trained multi-source separation model applied to simulated TDI data. We first describe the characteristics of the training dataset, including the astrophysical and instrumental components used to construct the input mixtures. We then examine the learned latent space representation, providing insight into how the shared encoder organizes different source types. Finally, the model's performance is evaluated across a range of test scenarios, demonstrating its ability to extract overlapping sources, reconstruct weak signals, and handle realistic noise conditions.

\subsection{Training data and simulation setup}

\nh{

The training dataset consists of simulated LISA TDI time series, incorporating a superposition of merging MBHBs, GBs, transient glitches, and stationary instrumental noise. The data is generated using the \texttt{BBHx} package for MBHBs \cite{Katz:2020hku,Katz:2021uax, michaelkatz_2021} and \texttt{FastGB} for GBs \cite{PhysRevD.76.083006}, ensuring physically motivated waveforms. Each time series is sampled at a rate of $\Delta t = 5$ seconds, with individual training snippets lasting for 2 hours.

\paragraph*{MBHBs.} The MBHB waveforms span component masses uniformly sampled in between \(10^5 M_\odot\) and \(10^6 M_\odot\), with redshifts drawn uniformly in comoving volume over the range \(z = 2\) to \(z = 5\). The binaries are assumed to be spin-aligned and non-precessing, and the waveforms include inspiral, merger, and ringdown phases with higher-order harmonics using the \texttt{IMRPhenomHM} approximant.

\paragraph*{GBs.} GB signals are modeled as slowly drifting sinusoids to reflect the intrinsic frequency evolution of compact white-dwarf binaries over multi-hour timescales. Their frequencies are drawn uniformly in log-space from 1\,mHz to 10\,mHz, and the strain amplitudes are sampled uniformly in log-space between \(1 \times 10^{-22}\) and \(2 \times 10^{-21}\). For each sample, up to five GBs are included (\(N_{\text{GB}} \leq 5\)), with the actual number per simulation drawn from a uniform discrete distribution. Sky positions are sampled isotropically, and polarization and inclination angles are drawn uniformly over their natural ranges. 

While this setup does not include the full Galactic foreground, which is expected to form a confusion-limited noise floor in LISA's low-frequency band, it enables us to assess the model's separation capabilities in interpretable conditions. We acknowledge that a realistic mission scenario will operate near the detection limit for most GBs as spectral resolution improves and that resolving overlapping near-threshold sources will be significantly more challenging. Scaling to such high-density GB populations is underway and will be addressed in future work.

\paragraph*{Frequency binning.} The frequency range from 1\,mHz to 10\,mHz is divided into \( K = 5 \) uniform-width bins, each spanning 1.8\,mHz. This bin count matches the maximum number of overlapping GB sources per training sample, providing a clean mapping in which each source can, in principle, be assigned to its own non-overlapping spectral bin. Within each bin, sources are aggregated and treated as a single target during training, and the model learns to reconstruct this frequency-binned representation. 

While this design simplifies the problem and enables a proof of concept, it does not reflect the full complexity expected in LISA data. In reality, GB sources will be densely distributed in frequency space, with many overlapping in narrow bands. The current setup thus represents a tractable starting point for demonstrating the feasibility of source separation in moderately crowded conditions. We emphasize that this binning strategy is a first step, and future work will address more realistic scenarios involving higher GB densities and stronger spectral overlap. To support this, we outlined architectural enhancements in the previous section.

\paragraph*{Glitches.}
Transient glitches are modeled as localized Gaussian bursts, where the amplitude is chosen to yield a broad range of SNRs. Specifically, we target glitches ranging from low-SNR cases that are buried in the quasi-stationary noise to high-SNR transients that can exceed the peak amplitude of the MBHB signal. Each sample includes between 0 and 30 such glitches, with randomly sampled locations. 

\paragraph*{Noise model.} The instrumental noise follows the standard LISA noise curve, including optical metrology and test mass acceleration noise contributions. It is illustrated in Fig.~\ref{fig:sensitivity-curve}.

\paragraph*{Training details.} We train the model using the Adam optimizer with a fixed learning rate of \(10^{-3}\), optimizing the loss function defined in Eq.~\ref{eq:TotalLoss}. Each model is trained for 25 epochs with a batch size of 16, ensuring convergence across all decoder heads. The training dataset contains 25{,}000 samples.

\begin{figure}[t!]
    \centering
    \includegraphics[width=0.46\textwidth]{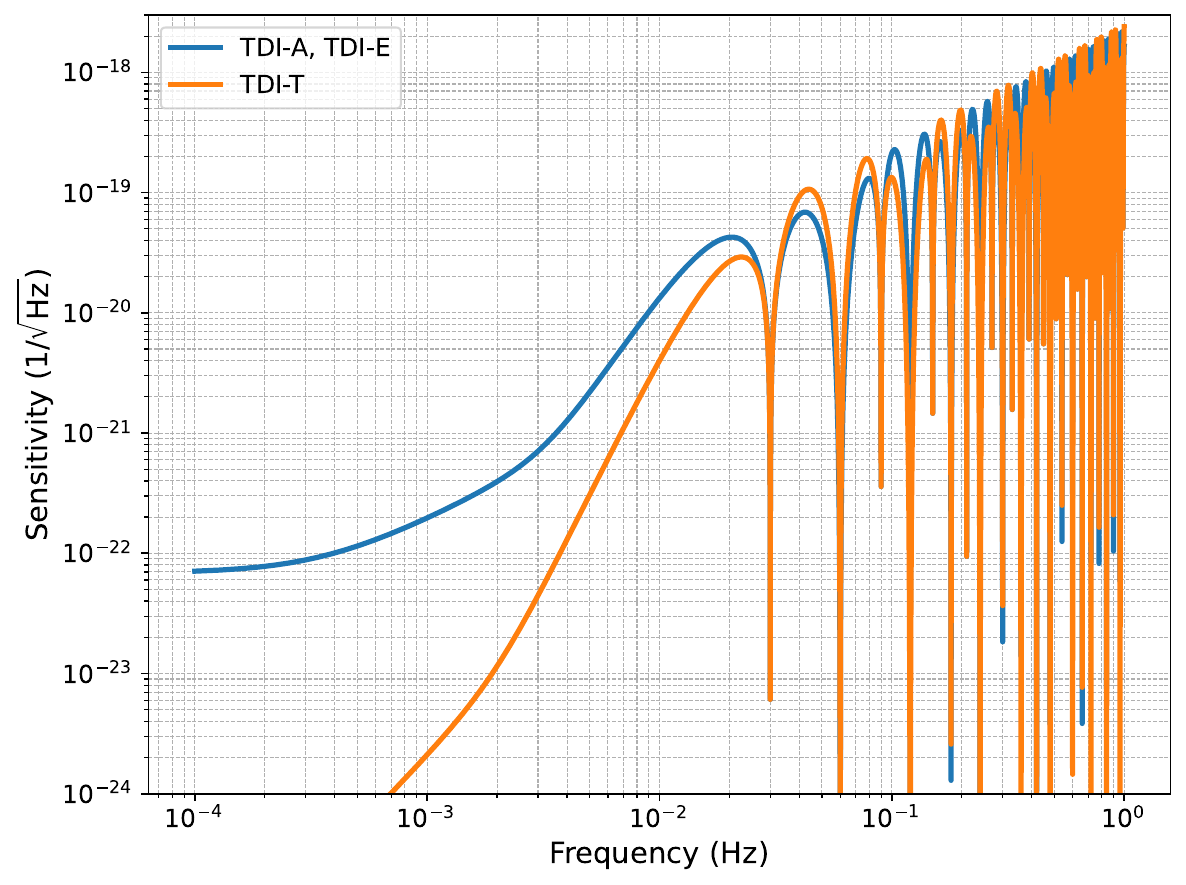}
    \caption{\RaggedRight LISA noise curves for the TDI A, E, and T channels. We use these profiles to generate colored Gaussian noise in our simulations, ensuring that the synthetic noise matches the expected characteristics of the LISA instrument.}
    \label{fig:sensitivity-curve}
\end{figure}
}

\begin{figure}[t!]
    \centering    \includegraphics[width=0.441\textwidth, trim=60 60 60 60, clip]{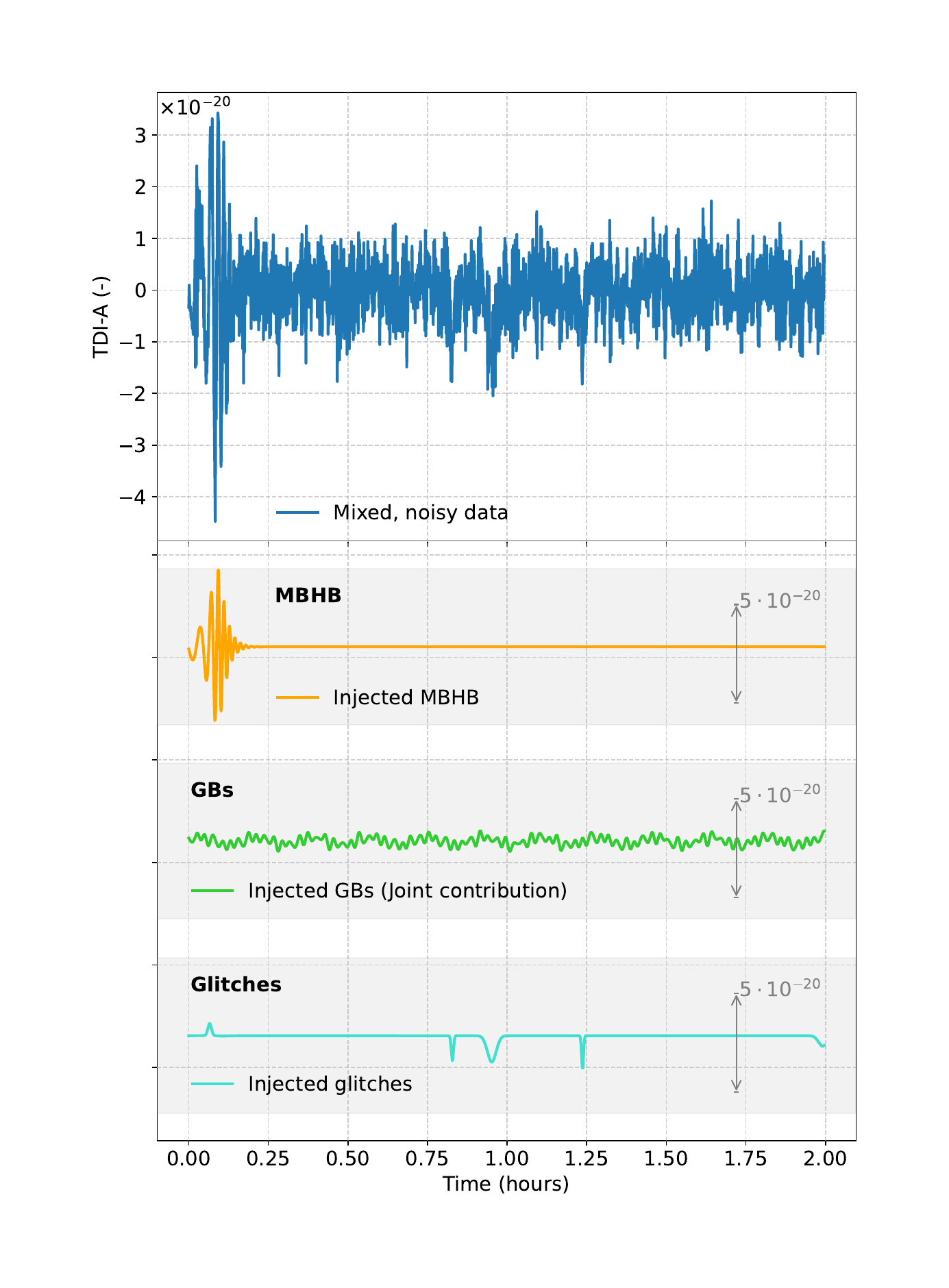}
    \caption{\RaggedRight Example of a time-domain representation of LISA's TDI data, capturing contributions from a merging MBHB, GBs, glitches, and stationary noise, used for model training and testing. The upper panel presents the noisy TDI-A channel, while the lower panels decompose it into individual signal components, which the proposed framework seeks to separate. Note that in this evaluation, we display solely the second-generation TDI-A channel, excluding the E and T observables.}
    \label{img:Example-Signal}
\end{figure}

Figure \ref{img:Example-Signal} presents a representative training sample, highlighting the interplay of MBHBs, GBs, glitches, and noise in the top panel, with the individual components displayed below. To start, the glitch distribution here is relatively simple. We will present more complex signal-artifact overlaps and glitch patterns throughout this section.

\vspace{-10pt}
\subsection{Latent space representation}

\begin{figure}[]
    \centering    \includegraphics[width=0.48\textwidth, trim=60 50 0 50, clip]{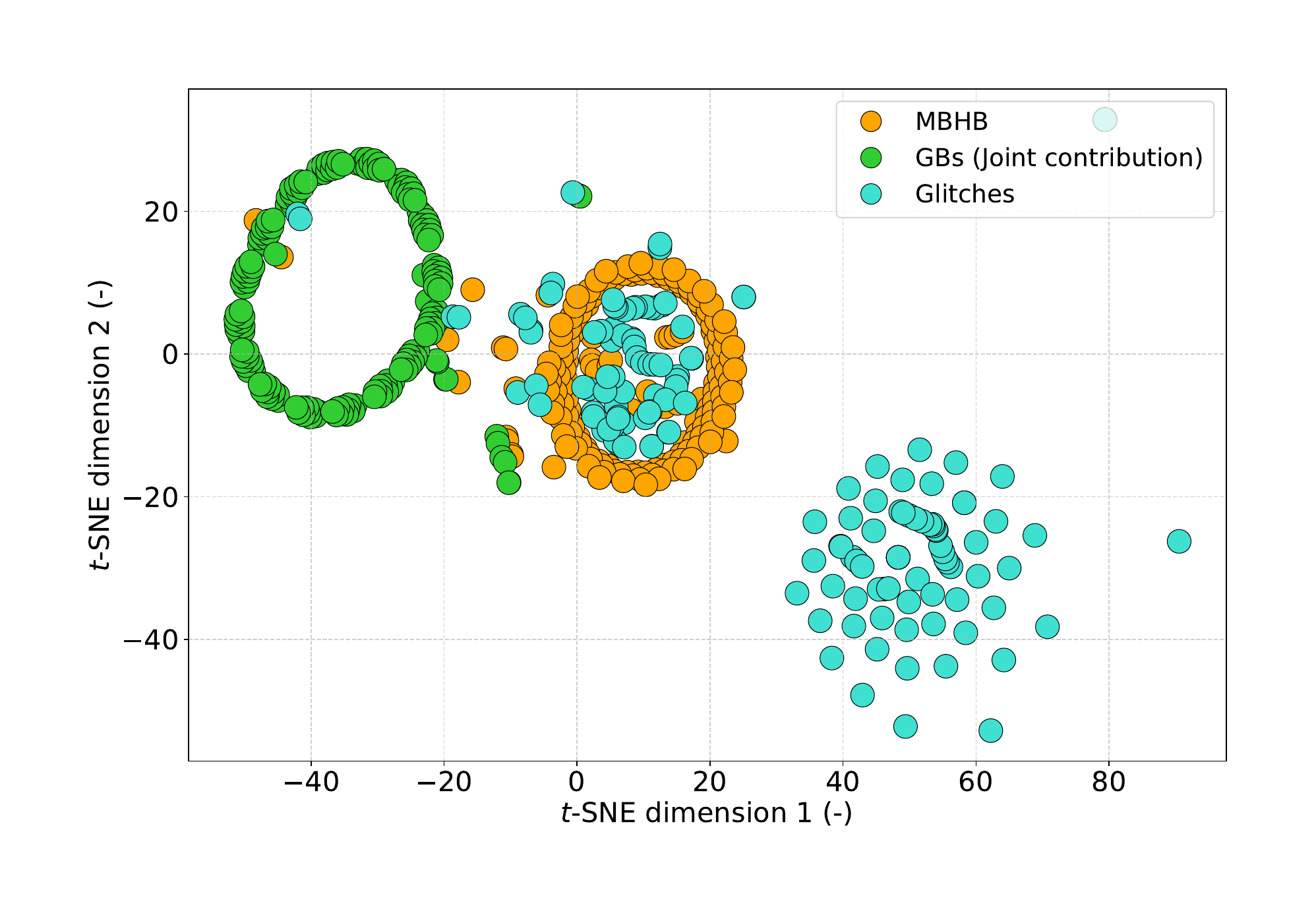}
    \caption{\RaggedRight $t$-SNE projection of bottleneck-encoded features derived from the normalized time series data in Fig.~\ref{img:Example-Signal}, illustrating clustering of merging MBHBs, GBs, and glitches. Note that separating sources within an abstract feature space beyond traditional temporal and spectral domains denotes a reorientation of methodology in LISA data analysis.}
    \label{img:Example-TSNE}
\end{figure}

To analyze the structure of the learned latent representation, we extract and visualize the bottleneck-encoded features from the trained model. This visualization provides insight into how different signal components are represented in the latent space. Figure~\ref{img:Example-TSNE} displays the two-dimensional $t$-SNE projection of the latent representations obtained from the encoded features of the time series in Fig.~\ref{img:Example-Signal}. Each component -- MBHB, GBs, and glitches -- is processed separately through the trained encoder and bottleneck layer, producing latent space representations of shape (256, 180), where 256 corresponds to the number of feature channels, and 180 represents the compressed temporal dimension. Before applying $t$-SNE, these latent features are reshaped and standardized using $z$-score normalization. Specifically, we concatenate the latent representations across the three signal types into a single dataset, ensuring that all features are on a comparable scale. We use the $t$-SNE algorithm \cite{JMLR:v23:21-0524} with a perplexity of 15. The perplexity parameter in $t$-SNE controls the balance between local and global structure in the projection. Lower perplexity values emphasize local structure, while higher values capture more global relationships. Each time step from the bottleneck layer is visualized as an individual data point in the scatter plot, color-coded by its corresponding source category.

The $t$-SNE projection reveals distinct clustering patterns associated with different signal components, indicating a degree of structure in the learned latent space. While $t$-SNE is a non-linear method that does not guarantee to preserve global distances, its ability to highlight local relationships allows us to identify grouping tendencies within the latent space. For instance, the MBHB (orange) and GB (green) components display distinct ring-like structures. While these patterns may be partially influenced by the \textit{crowding problem}, where high-dimensional data is compressed into a lower-dimensional representation, their presence suggests that the model has learned to encode different signal types in a structured manner. Such clustering behavior, even if influenced by the properties of $t$-SNE, reflects an underlying organization in the learned representations.

Glitches (turquoise) appear more dispersed, forming multiple clusters with some points scattered throughout the latent space. This dispersion may indicate challenges in encoding glitches into a single representation but could also reflect a genuine structural variation in the data. Some overlap between glitch and astrophysical signal clusters suggests possible feature entanglement, which may be mitigated through further refinements to the model architecture or loss function.

Overall, the observed clustering in the $t$-SNE projection supports the idea that the model has captured structured representations of the data. While $t$-SNE does not provide definitive proof of disentanglement due to its tendency to distort global relationships, it offers valuable insight into the organization of latent features. To further validate the model’s representation learning, complementary dimensionality reduction techniques such as Principal Component Analysis (PCA) \cite{Pearson1901} or Uniform Manifold Approximation and Projection for Dimension Reduction (UMAP) \cite{McInnes2018} could be applied. A more rigorous evaluation, namely direct signal reconstruction quality, is performed in the following to confirm that the representation of TDI data in a latent space is indeed useful for our practical application.

\subsection{Model evaluation on unseen test data}

To evaluate the model's generalization capability, we apply it to unseen test data and compare its predictions to ground-truth signals. 

\begin{figure}[]
    \centering    \includegraphics[width=0.441\textwidth, trim=60 60 60 60, clip]{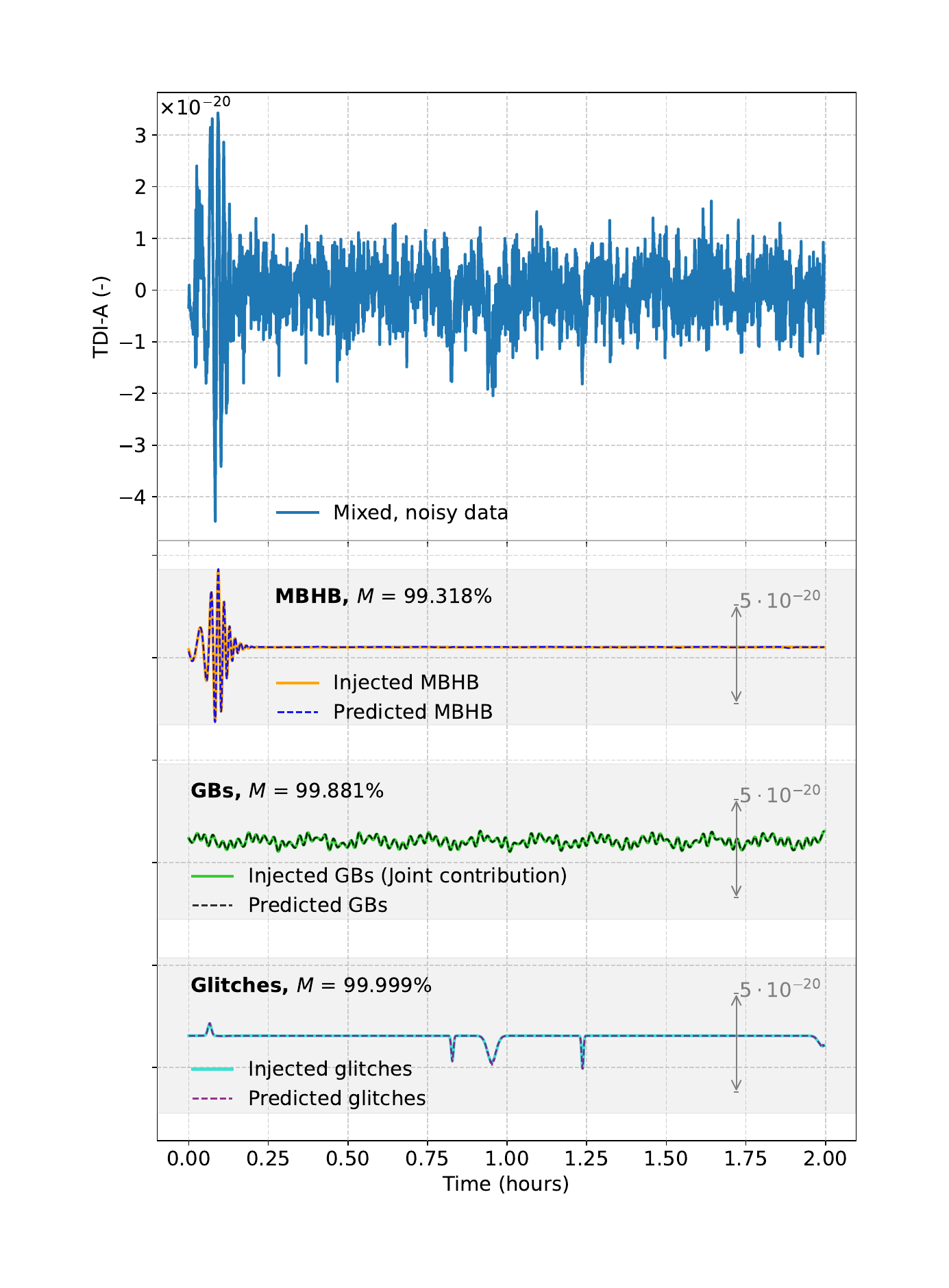}
    \caption{\RaggedRight Time-domain predicted waveforms from the deep source separation model overlaid on the true waveforms from Fig.~\ref{img:Example-Signal}, illustrating the model’s ability to accurately disentangle, reconstruct, and de-noise individual components.}
    \label{img:Example-SignalPrediction}
\end{figure}

Figure \ref{img:Example-SignalPrediction} shows the reconstructed MBHBs, GBs, and glitches of Fig.~\ref{img:Example-Signal} obtained by the decoder heads and the bottleneck features visualized in Fig.~\ref{img:Example-TSNE}. The upper panel illustrates the signal mixture, which serves as the input for our framework. The lower panels display the contributions separated for clarity, where the injected waveforms are compared with the corresponding estimates obtained with the deep source separation approach. 

To validate the quality of the learned encoder-decoder framework, we use the absolute normalized match factor, which quantifies the similarity between a predicted waveform $\hat{x}$ and a true waveform $x$ on TDI level. The metric is defined as
\begin{equation}
M = \frac{| \langle x | \hat{x} \rangle |}{\sqrt{\langle x | x \rangle \langle \hat{x} | \hat{x} \rangle}},
\end{equation}
where the inner product $\langle x | \hat{x} \rangle$ is weighted by the noise power spectral density $S^X_n(f)$ in a given TDI channel:
\begin{equation}
\langle x | \hat{x} \rangle = \sum_{f} \frac{X(f) \hat{X}^*(f)}{S^X_n(f)}.
\end{equation}
Here, $X(f)$ and $\hat{X}(f)$ are the Fourier transforms of $x$ and $\hat{x}$, respectively. This noise-weighted inner product is standard in gravitational-wave data analysis and reflects the optimal matched filtering statistic under Gaussian noise assumptions. The weighting by $1/S_n^X(f)$ down weights frequency regions with high noise (low sensitivity) and emphasizes those where the detector is most sensitive. As such, it ensures that waveform agreement is judged in terms of physically relevant distinguishability in the presence of instrumental noise.

The decoders accurately recover the merging MBHB, glitches, and GBs signals for this simple example. Regarding the multi-output channel GB decoder, we present only the prediction of the joint GB contribution obtained by summing all individual GB predictions. An analysis of the individually resolved GB sources follows at the end of this section. We further examine the model's robustness by analyzing specific test cases that challenge its separation ability. These scenarios include strong overlaps between glitches and MBHBs, weak MBHB signals buried in noise, quiet periods without MBHBs, and high-density GB regions.

\subsubsection{Glitches overlapping with MBHB signal} 

\begin{figure*}[]
    \centering
    \begin{subfigure}{0.44\textwidth}
        \centering
        \includegraphics[width=0.95\textwidth, trim=0 0 0 0, clip]{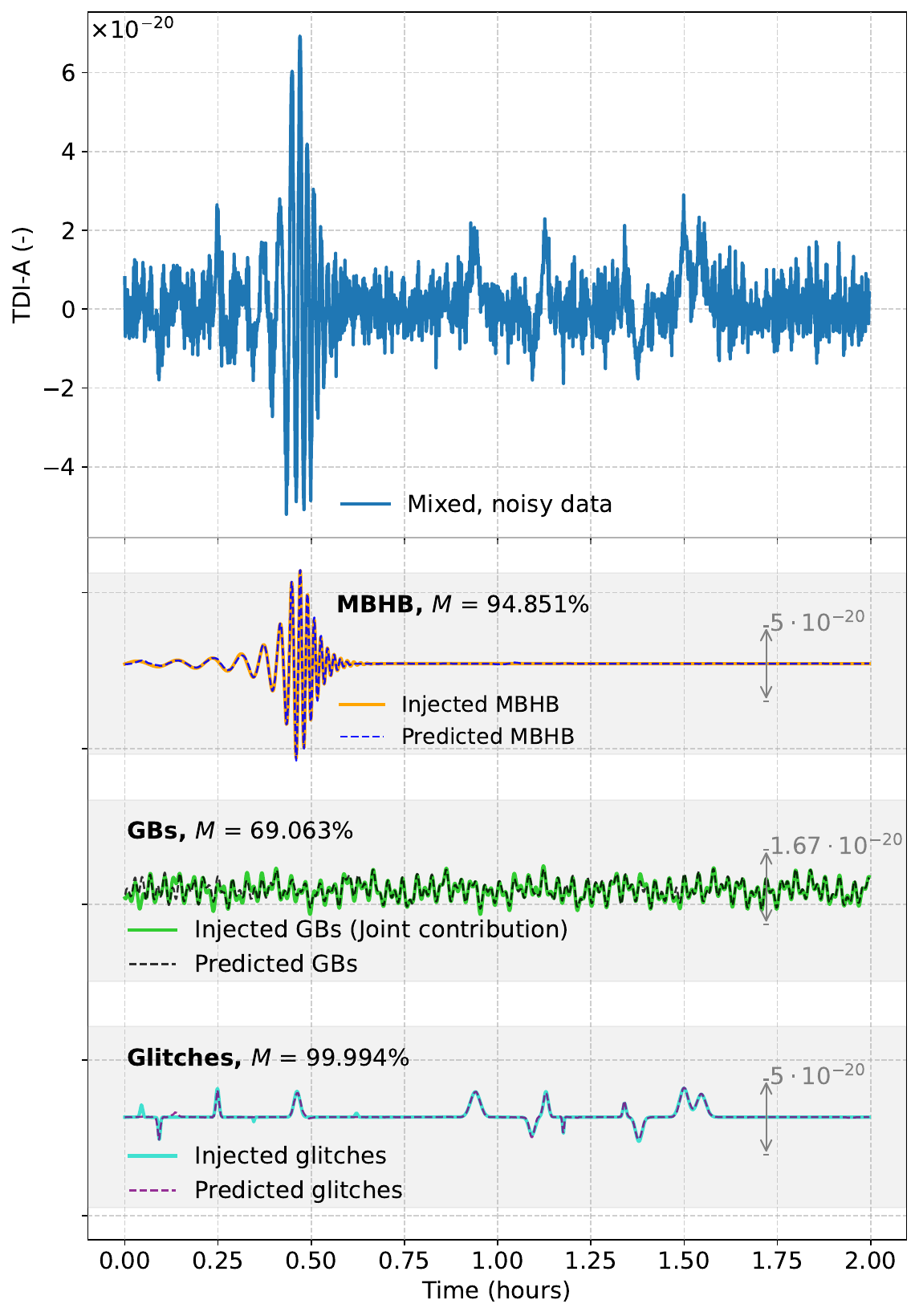}
        \caption{}
    \end{subfigure}
    \hfill
    \begin{subfigure}{0.44\textwidth}
        \centering
        \includegraphics[width=0.95\textwidth, trim=0 0 0 0, clip]{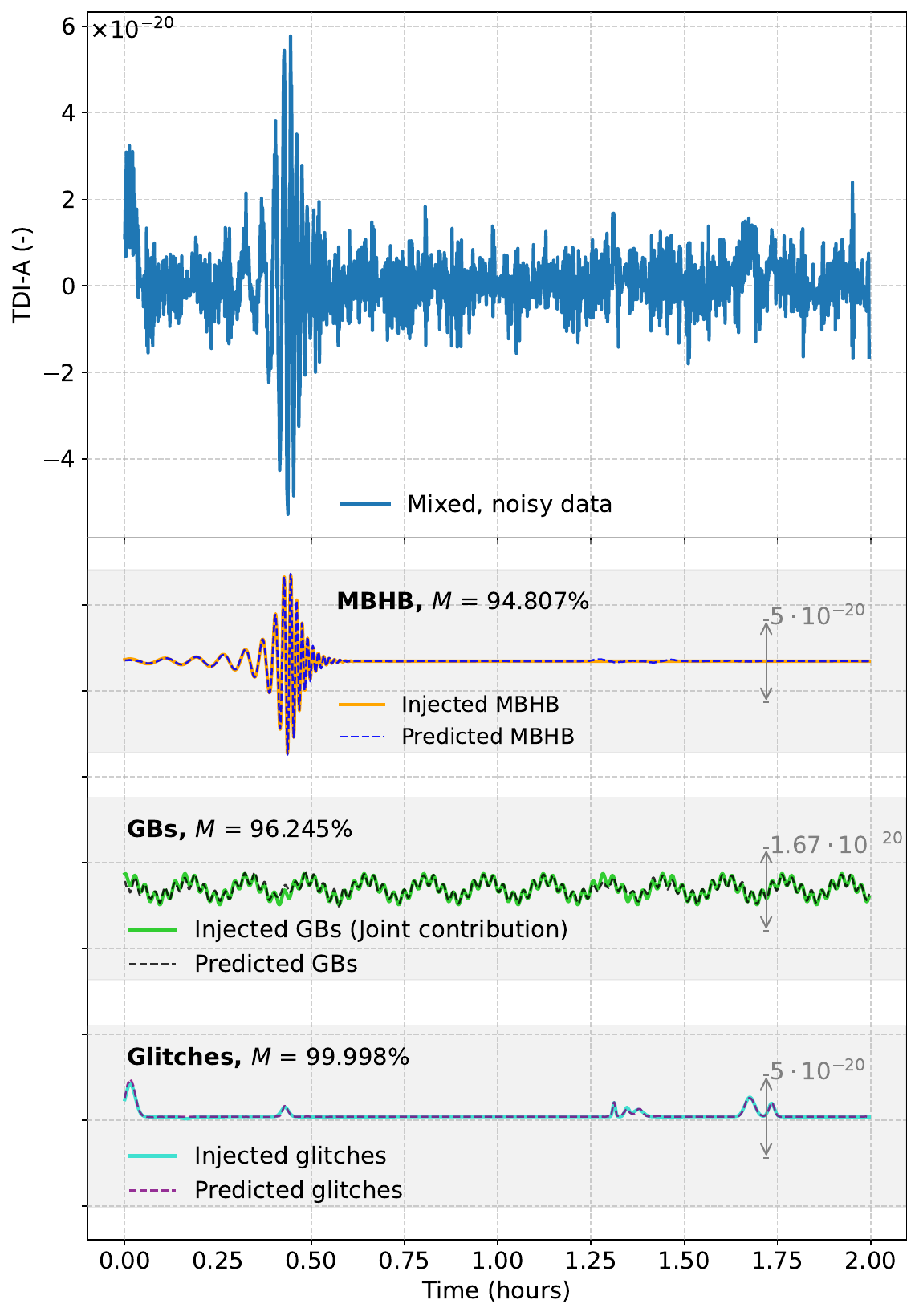}
        \caption{}
    \end{subfigure}
    
    \vspace{0.1cm} 
    
    \begin{subfigure}{0.44\textwidth}
        \centering
        \includegraphics[width=0.95\textwidth, trim=0 0 0 0, clip]{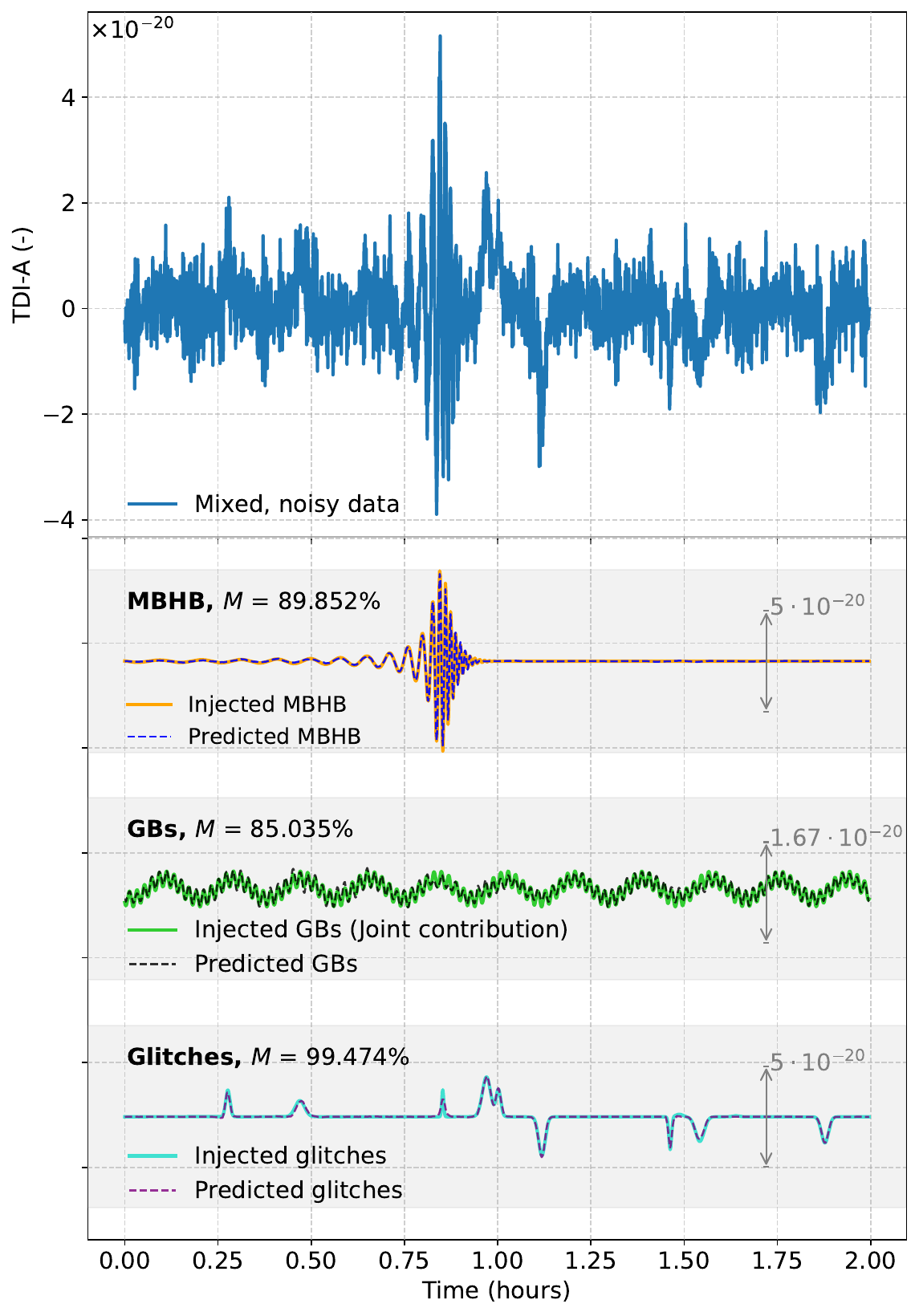}
        \caption{}
    \end{subfigure}
    \hfill
    \begin{subfigure}{0.44\textwidth}
        \centering
        \includegraphics[width=0.95\textwidth, trim=0 0 0 0, clip]{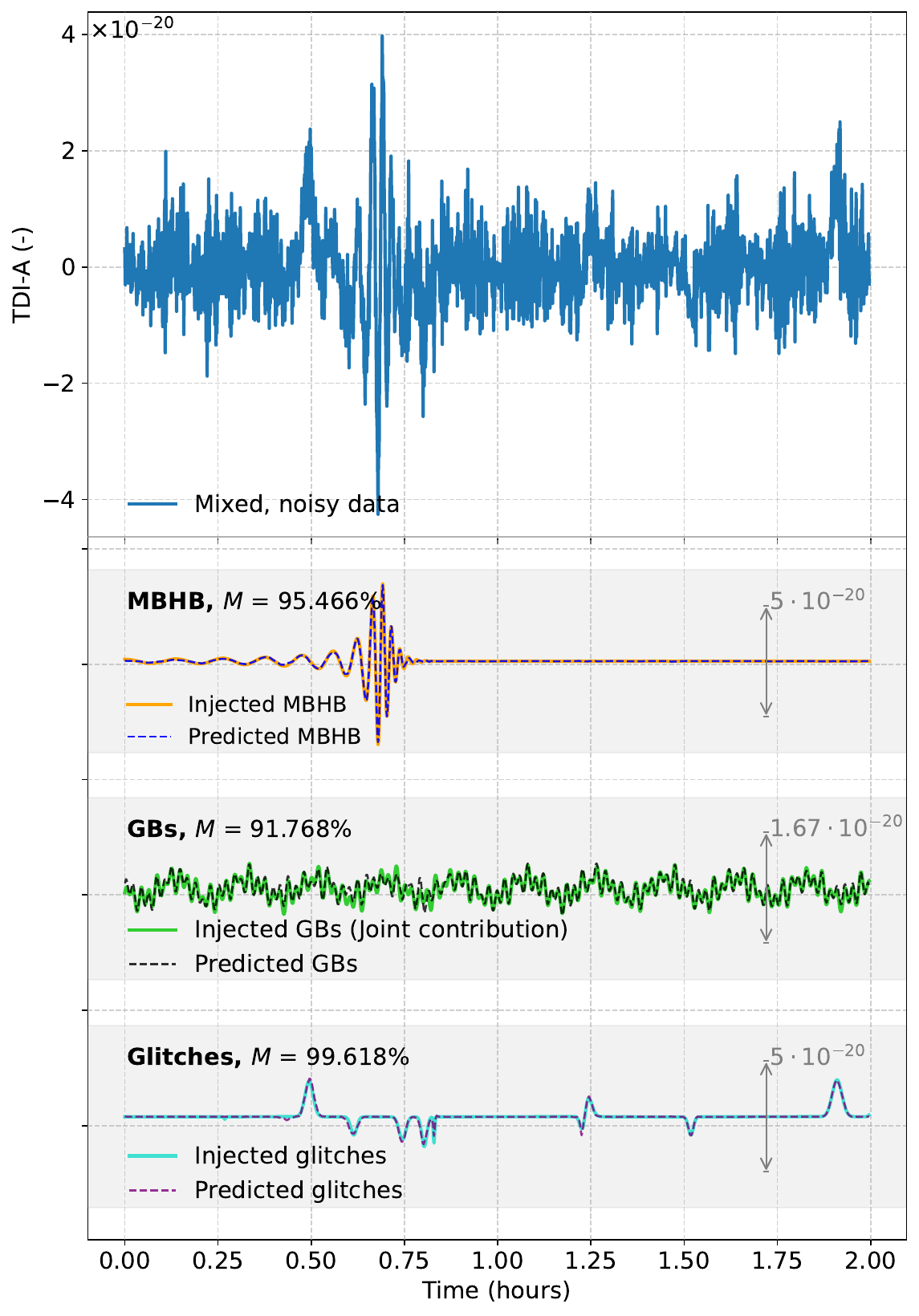}
        \caption{}
    \end{subfigure}
    
    \caption{\RaggedRight Injected waveforms and model predictions in the presence of overlapping instrumental glitches. Panels (a) and (b) show cases where glitches overlap with the MBHB merger phase, while (c) and (d) illustrate glitches occurring during the MBHB ringdown.}
    \label{fig:SignalPrediction_Overlap}
\end{figure*}

A critical challenge in LISA data analysis is distinguishing instrumental glitches from astrophysical sources. To test the model's performance in such cases, we consider examples where glitches overlap with MBHB waveforms during their inspiral, merger, and ringdown phases. Figure \ref{fig:SignalPrediction_Overlap} compares the raw input signal, the true waveforms, and the model's predicted outputs in such scenarios.

\nh{To further quantify the impact of glitches on MBHB waveform recovery, we simulate datasets that include an MBHB signal, a single instrumental glitch, and stationary LISA-like noise. In each simulation, the MBHB and glitch components are independently normalized to achieve comparable SNRs, ensuring that both contribute similarly to the time-domain mixture.
Balancing the SNRs establishes a controlled setting where both the astrophysical signal and the glitch influence the data similarly, preventing trivial cases in which one component dominates. This setup allows us to probe their interference and disentanglement during recovery meaningfully.

We then systematically vary the relative timing between the MBHB coalescence and the glitch, shifting the glitch by an offset ranging from –2 hours to 0 (coalescence time) while keeping the MBHB and noise fixed. At each offset, we generate the noisy TDI mixture, apply the deep source separation model, and compute the noise-weighted match factor between the true and recovered MBHB waveforms.
Figure \ref{fig:Match_vs_GlitchOffset} displays the median match factor $M$ as a function of glitch offset (solid red line), along with the 25–75\% interquartile range (dark shaded region) and the 5–95\% percentile range (light shaded region) over 150,000 randomized glitch–MBHB realizations. The x-axis denotes the glitch offset in hours before coalescence, with the vertical dashed line marking the moment of the MBHB merger.

\begin{figure}[htbp]
    \centering
     \hspace*{-1cm} 
    \includegraphics[clip, trim=0 0 0 0, width=0.5\textwidth]{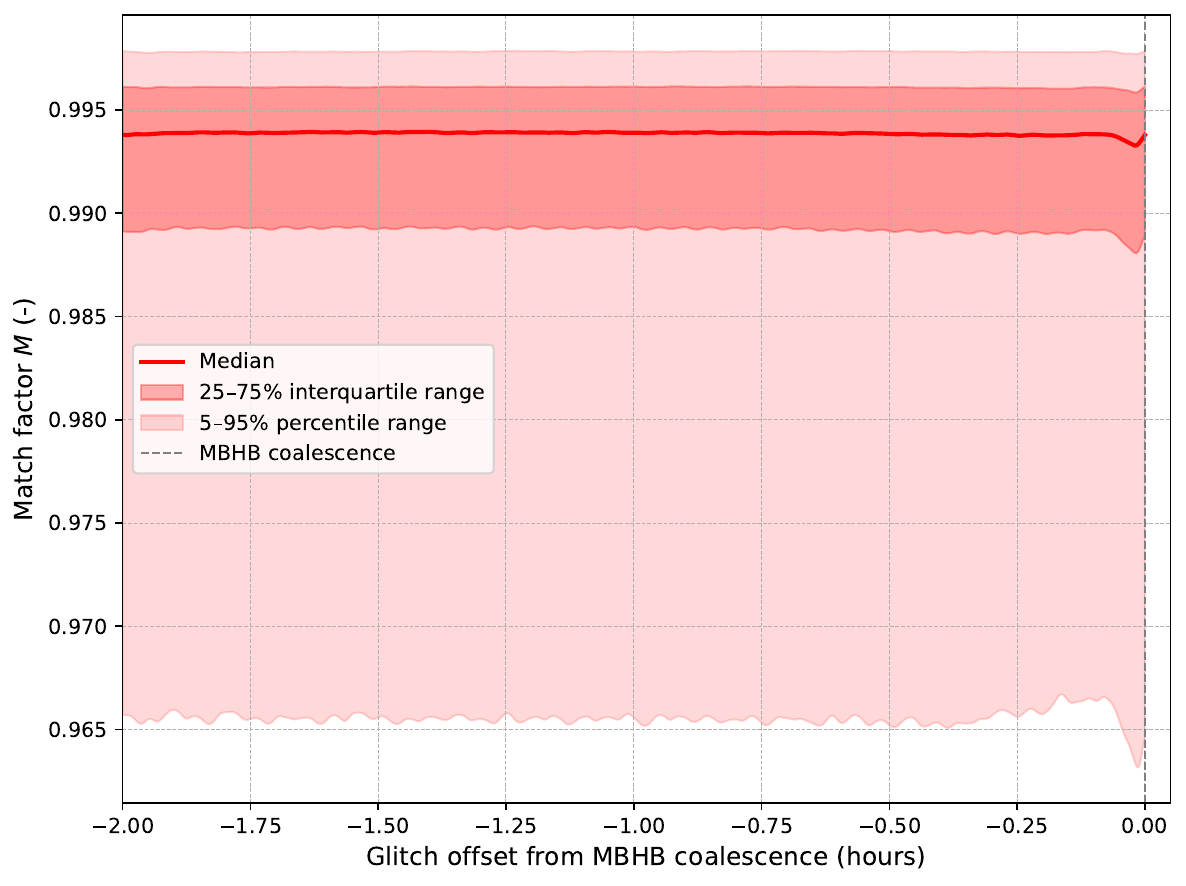}
    \caption{\RaggedRight
      Impact of glitch timing on MBHB recovery. We simulate mixtures of an MBHB signal, a single glitch, and LISA-like noise, scaling the MBHB and glitch to similar SNRs. In this figure, we set the SNR to 50. The glitch is systematically shifted in time relative to the MBHB coalescence, and the deep source separation model is used to recover the MBHB waveform across all offset configurations. The match factor remains nearly constant, with a slight decline only when the glitch overlaps the merger. Higher SNRs yield higher match values, but the overall behavior remains consistent.
    }
    \label{fig:Match_vs_GlitchOffset}
\end{figure}

The match factor remains consistently high across most of the inspiral phase, with only a modest decline occurring when the glitch temporally overlaps with the merger. This suggests that the deep source separation model can quite robustly recover MBHB signals. One reason for this robustness lies in the distinct spectral and temporal characteristics of MBHB signals versus glitches. MBHB mergers are coherent chirps, whereas glitches may appear as abrupt, high-frequency bursts or narrow-band transients. The MBHB decoder learns to recognize the typical MBHB waveform morphology in the latent space representations, even in time-overlapping cases.
}
\vspace{-5pt}
\subsubsection{Weak MBHB mergers buried in noise} 
\begin{figure*}
    \centering
    \begin{subfigure}{0.45\textwidth}
        \centering
        \includegraphics[width=0.98\textwidth, trim=0 0 0 0, clip]{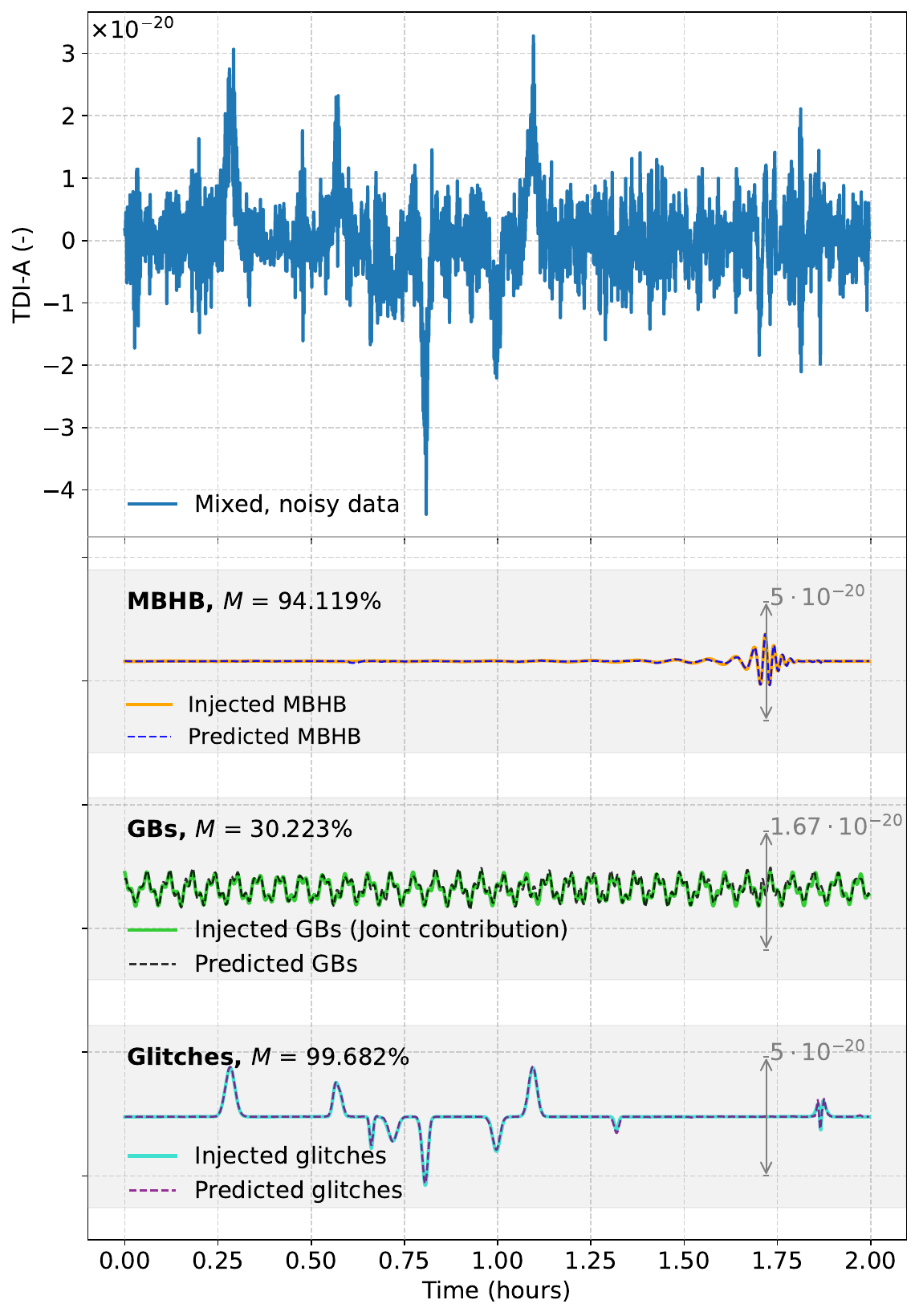}
        \caption{}
    \end{subfigure}
    \hfill
    \begin{subfigure}{0.45\textwidth}
        \centering
        \includegraphics[width=0.98\textwidth, trim=0 0 0 0, clip]{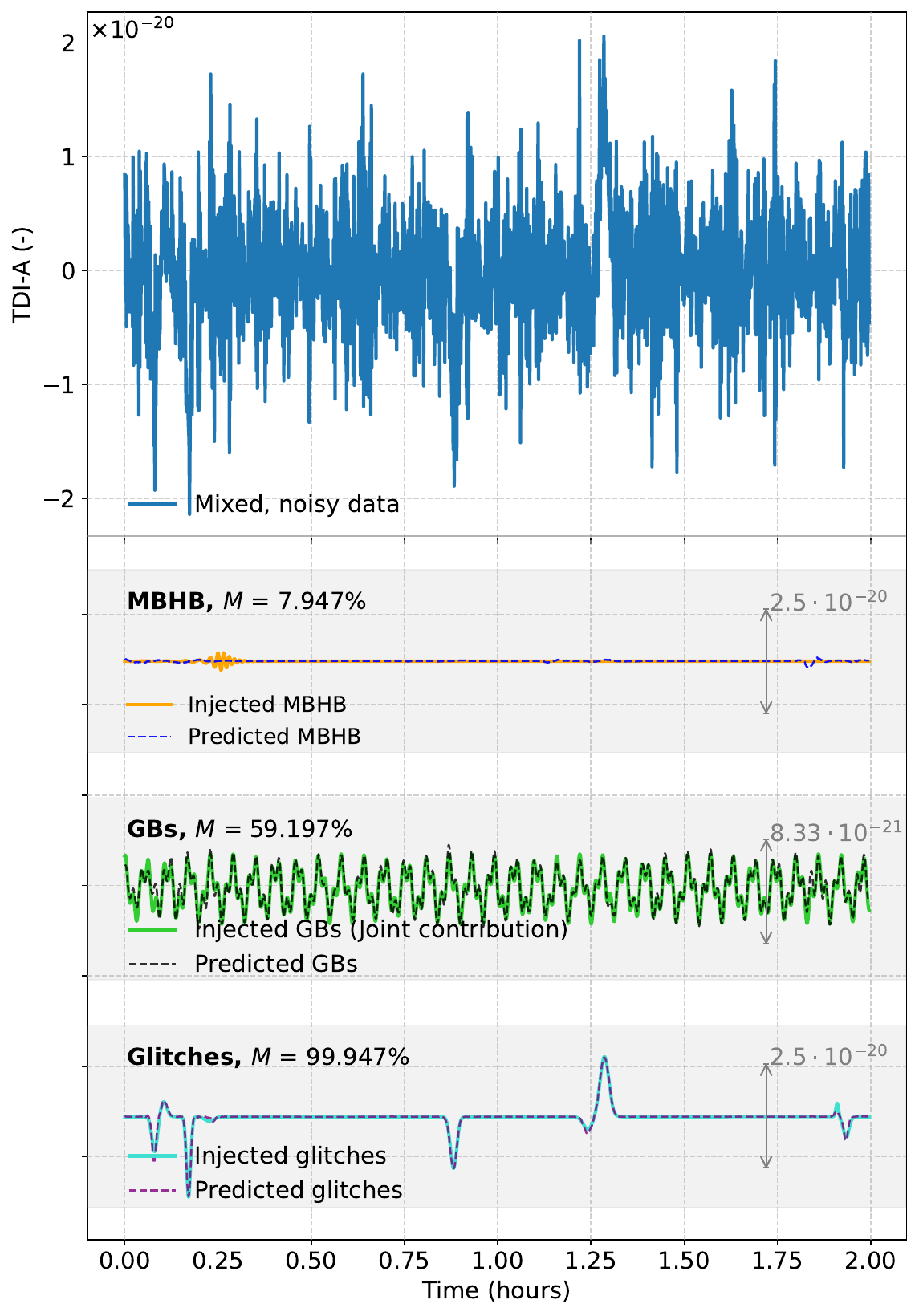}
        \caption{}
    \end{subfigure}
    \caption{\RaggedRight Comparison of injected waveforms and model predictions for a low-amplitude MBHB merger buried in stationary noise. In panel (a), the deep source separation framework successfully detects and reconstructs the MBHB signal. However, in (b), where the signal amplitude is further diminished, the model fails. In such cases, further investigation is needed to determine whether the issue lies in the shared encoder or the MBHB decoder head.}
    \label{img:Example-SignalPrediction_Weak}
\end{figure*}

\vspace{-3pt}
Another important test is the model's ability to extract weak MBHB signals from the noise floor. In this scenario, the MBHB is barely visible in the input time series, simulating the detection of high-redshift mergers. Figure \ref{img:Example-SignalPrediction_Weak} presents two examples: one where the model successfully recovers the MBHB waveform, albeit with some power leakage into the glitch decoder, and another where it fails as the merger amplitude is further reduced.

\nh{Figure \ref{fig:match_vs_snr} illustrates the relationship between the SNR and the normalized match factor $M$ for MBHB signals recovered from noisy TDI data using the trained deep source separation model. Each gray point represents a single simulation; in total, 150{,}000 points are shown. The red curves indicate the median match (solid line), the 25–75\% interquartile range (dark shaded region), and the 10–95\% percentile spread (light shaded region).A vertical line marks the SNR at which the median match first exceeds the accuracy threshold of $M = 0.95$, which in this analysis occurs at approximately SNR $\approx 15$. This threshold is chosen for illustrative purposes. A systematic investigation is needed to determine how well waveform parameters can be recovered as a function of the match factor. We present preliminary results on this question at the end of this section, though we note that the primary focus of this paper is on deep source separation rather than source parameter inference.

\begin{figure}[]
    \centering
     \hspace*{-1cm} 
    \includegraphics[clip, trim=0 0 0 0, width=0.5\textwidth]{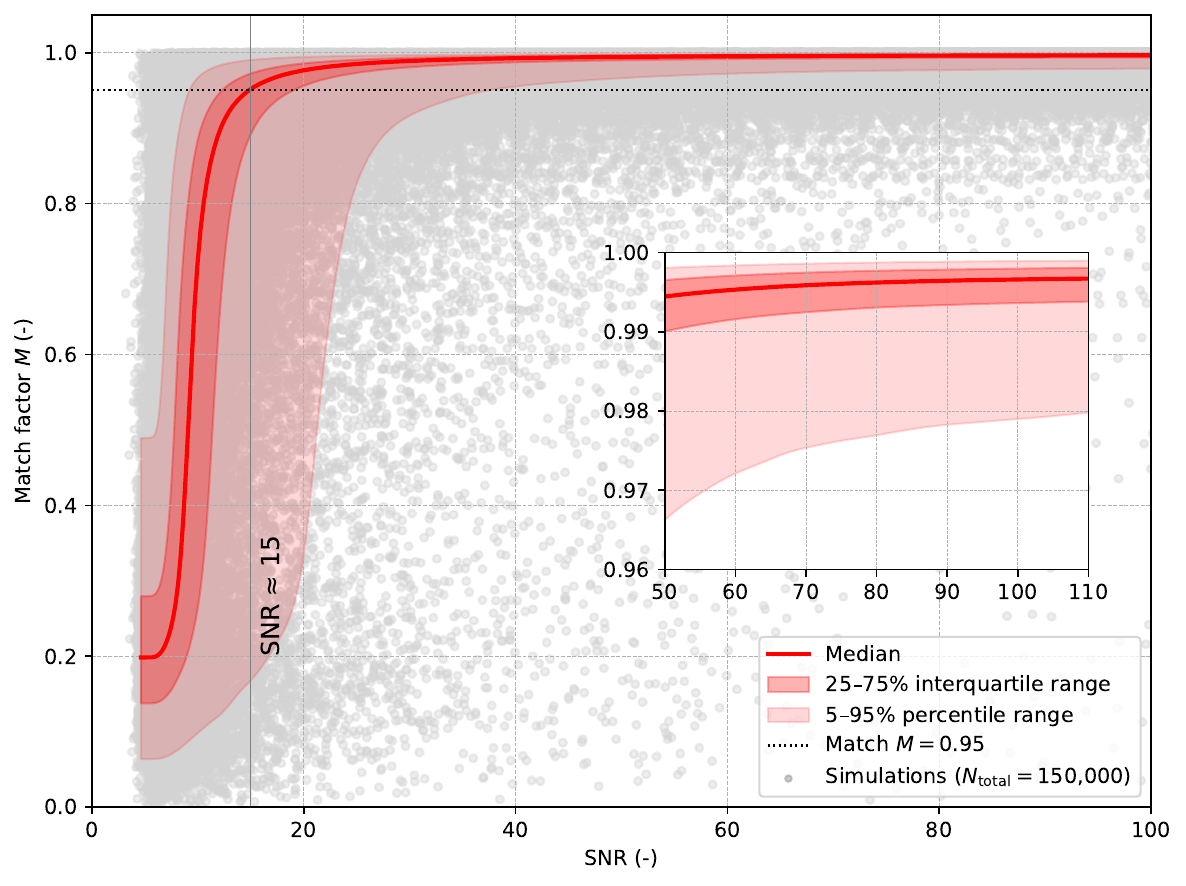}
    \caption{\RaggedRight
        Match factor $M$ as a function of SNR for quiet MBHB signals recovered from noisy TDI data. Gray points show individual simulations, and red curves indicate the median and percentile bands. The vertical line marks where the median first exceeds $M = 0.95$. The value is arbitrarily chosen for visualization. SNRs are computed empirically, treating glitches and GBs as noise.
    }
    \label{fig:match_vs_snr}
\end{figure}

To probe match performance across a broad SNR range, we generate multiple amplitude-scaled MBHB signals per realization, resulting in effective SNRs spanning from 5 to 100. In this setup, GBs and transient glitches are treated as part of the effective noise background, as they hinder the accurate recovery of the MBHB waveform. The empirical SNR for each signal is calculated by estimating the PSD from the effective noise background using Welch’s method.  The inset zooms into the high-SNR regime. The tightening of percentile bands with increasing SNR indicates improved reconstruction stability, while the saturation of the median match near unity confirms the model’s effectiveness in extracting MBHB signals under favorable conditions.}

\subsubsection{Quiet periods with no MBHB mergers} \vspace{-9pt}

\begin{figure}
        \centering
        \includegraphics[width=0.441\textwidth, trim=0 0 0 0, clip]{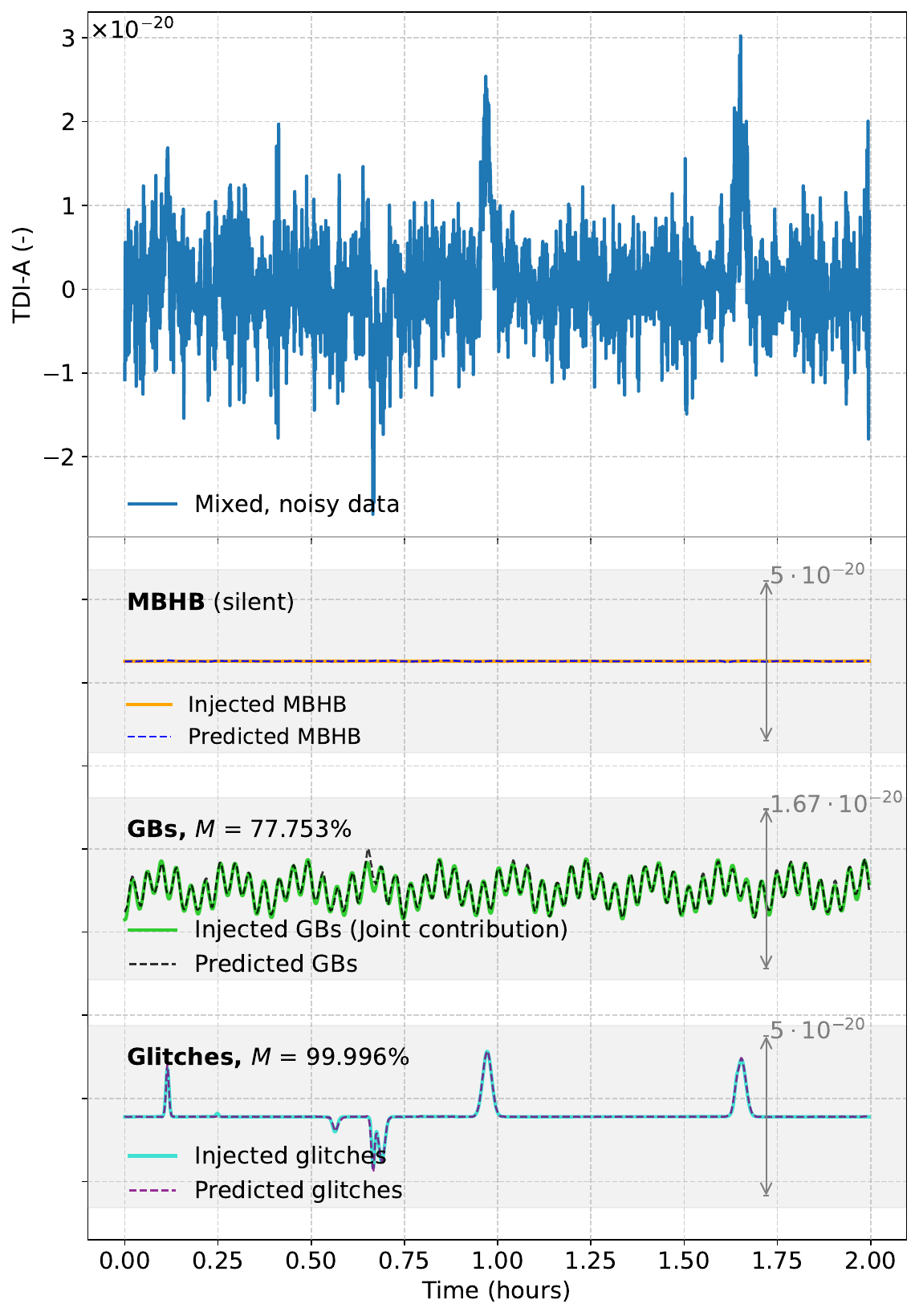}
        \caption{\RaggedRight Evaluation of the model in a time segment without a dominant MBHB to assess its ability to avoid false positives.}
        \label{img:Example-SignalPrediction_1929}
\end{figure}

An essential test for avoiding false positives is evaluating the model in time periods where no MBHB is present. Figure \ref{img:Example-SignalPrediction_1929} shows an input segment containing only GBs and stationary noise. The MBHB decoder output remains close to zero, indicating that the model does not hallucinate signals when none are present.

\nh{To further evaluate the reliability of the MBHB decoder in distinguishing true astrophysical signals from spurious activations, we analyze the empirical cumulative distribution functions (ECDFs) of the decoder output power. Specifically, we consider the squared $\ell_2$-norm of the decoder's output in the MBHB channel,
\begin{equation}
P = \sum_t  \hat{x}^2_{\mathrm{MBHB}}(t),
\end{equation}
where $\hat{x}_{\mathrm{MBHB}}(t)$ is the predicted strain at time $t$. The ECDF, defined as
\begin{equation}
\mathrm{ECDF}(P_0) = \frac{1}{N} \sum_{i=1}^{N} \mathbf{1}_{\{P_i \leq P_0\}},
\end{equation}
provides the cumulative fraction of samples whose decoder output power does not exceed a given threshold $P_0$. Intuitively, the ECDF tells us {for any power value, what fraction of decoder outputs were smaller than or equal to that value}.

We compare ECDFs for two types of simulated LISA data: one containing only instrument noise and glitches ("No MBHB"), and one in which a MBHB signal is injected ("With MBHB"). The gray curve in Fig.~\ref{fig:ECDF_MBHBs} shows the ECDF of decoder power in the absence of a true signal. As expected, the curve rises steeply and saturates at very low power levels, indicating that the decoder remains largely inactive when no MBHB is present. The red curve shows the ECDF for the same decoder when an MBHB signal is included, resulting in a markedly slower rise toward unity and significantly higher output power, reflecting strong decoder activation in response to astrophysical signals. In this case, we again use  150{,}000 simulation samples, with MBHB parameters randomly varied across the simulations.

\begin{figure}
    \centering
     \hspace*{-1cm} 
    \includegraphics[clip, trim=0 0 0 0, width=0.5\textwidth]{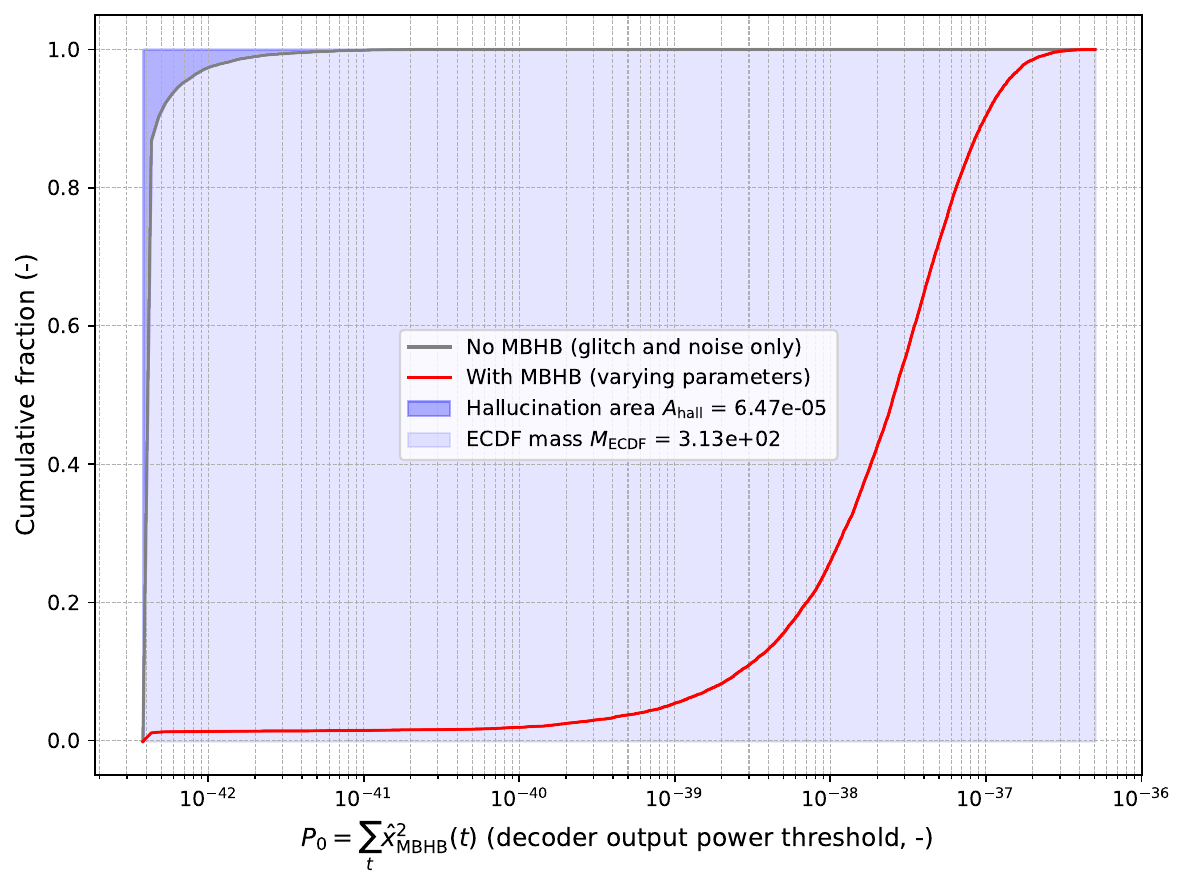}
    \caption{\RaggedRight
        ECDFs of the MBHB decoder output power, defined as the squared $\ell_2$-norm of the predicted strain. The gray curve shows results for glitch+noise-only inputs, while the red curve includes MBHB signals with varying parameters across simulations. The steep rise of the gray curve indicates minimal decoder activity during quiet periods. The small hallucination area and large ECDF mass confirm the decoder's low false-positive rate.
    }
    \label{fig:ECDF_MBHBs}
\end{figure}

To quantify potential spurious responses, we define the \emph{hallucination area} as the area between the actual ECDF for glitch+noise and the ideal ECDF that would jump directly to one at $P = 0$:
\begin{equation}
A_{\mathrm{hall}} = \int_{0}^{P_{\max}} \left( 1 - \mathrm{ECDF}_{\mathrm{quiet}}(P) \right) \, dP.
\end{equation}
This area captures the total "excess activation" of the decoder during quiet periods. A small hallucination area implies that the decoder rarely outputs significant power when no MBHB is present, indicating low false-positive risk. In our test data, we find $A_{\mathrm{hall}} = 6.88 \times 10^{-5}$.

In contrast, we define the \emph{ECDF mass} as the area under the ECDF curve itself:
\begin{equation}
M_{\mathrm{ECDF}} = \int_{0}^{P_{\max}} \mathrm{ECDF}_{\mathrm{quiet}}(P) \, dP,
\end{equation}
which reflects how quickly the decoder output accumulates across the dataset. A high ECDF mass corresponds to decoder outputs clustering near zero -- the ideal behavior when no MBHB is present. For the glitch+noise case shown, the ECDF mass is approximately $2.83 \times 10^2$.

Together, the separation between the red and gray curves, the small hallucination area, and the high ECDF mass all indicate that the MBHB decoder is well-calibrated and reliably silent during quiet periods while remaining sensitive to true astrophysical signals.}

\subsubsection{Resolving individual GBs in the presence of MBHBs and glitches}  \vspace{-9pt}

\begin{figure*}
    \centering
    \begin{subfigure}{0.45\textwidth}
        \centering
        \includegraphics[width=\textwidth, trim=0 0 0 0, clip]{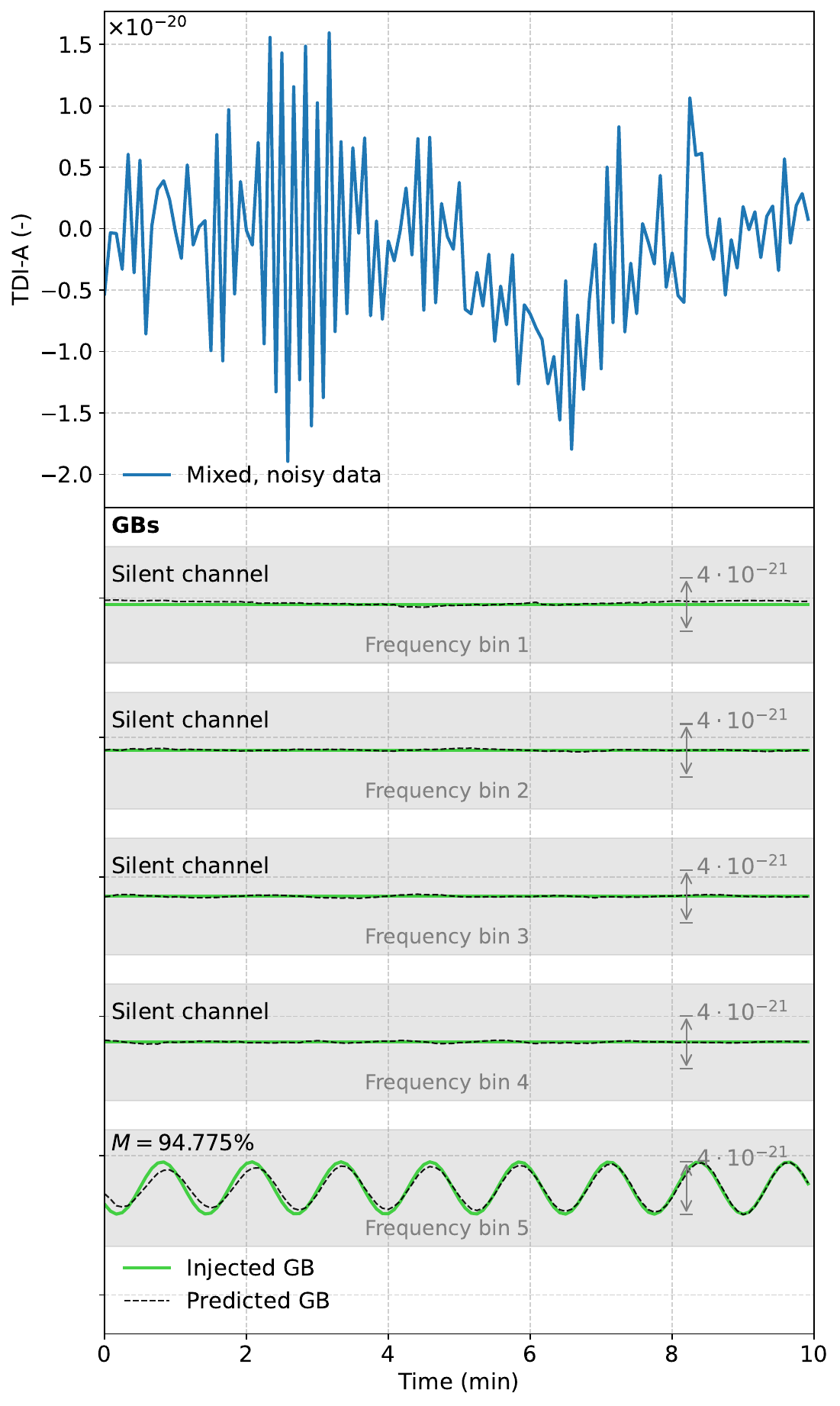}
        \caption{}
    \end{subfigure}
    \hfill
    \begin{subfigure}{0.45\textwidth}
        \centering
        \includegraphics[width=\textwidth, trim=0 0 0 0, clip]{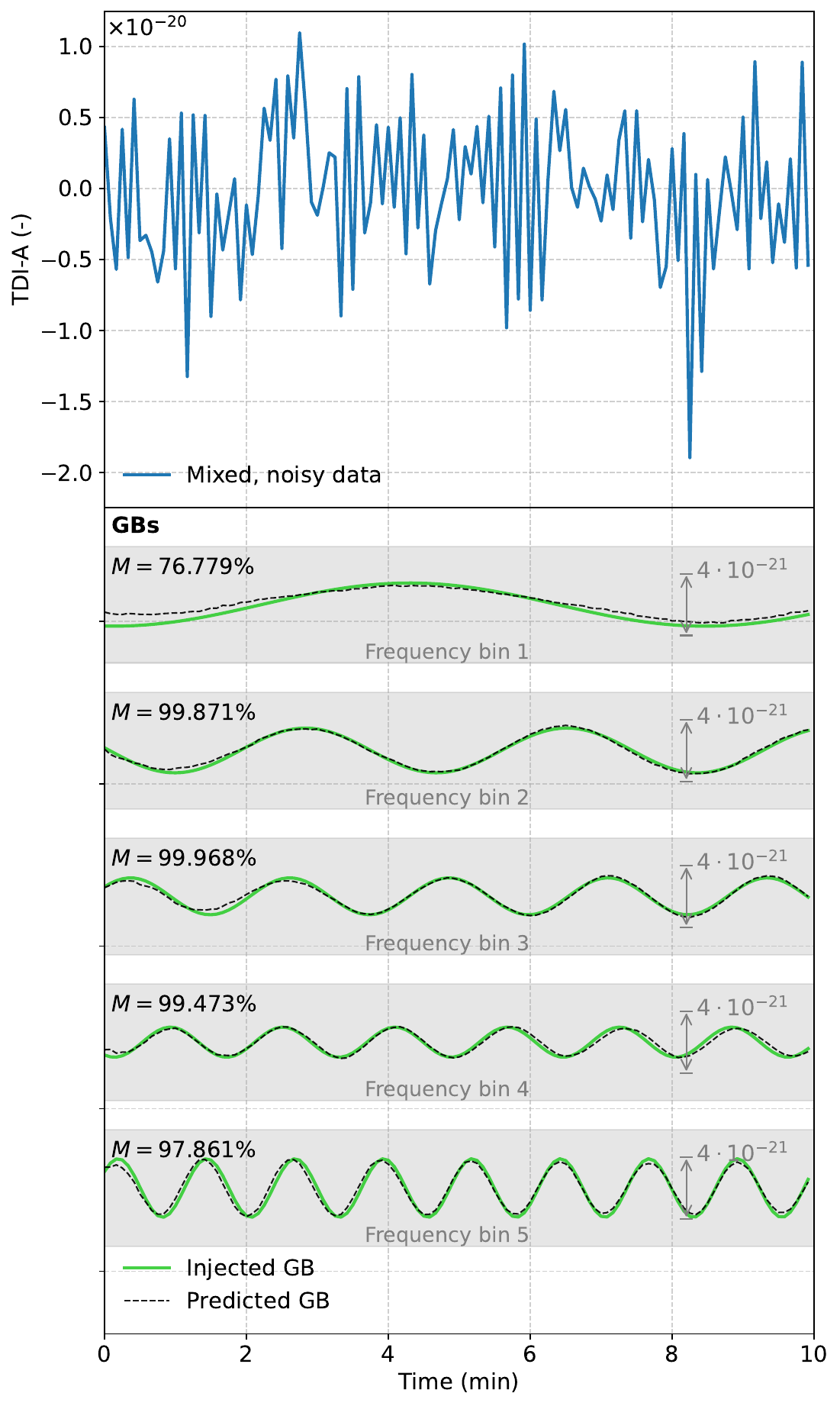}
        \caption{}
    \end{subfigure}
    \caption{\RaggedRight Comparison of the injected and predicted GB waveforms in the presence of glitches. We do not highlight the injected glitches explicitly in the noisy input dataset. The multi-channel GB decoder accurately estimates the number of active GB sources, i.e., one in panel (a) and five in (b). Scaling to the thousands of overlapping GBs expected in LISA will require larger training datasets and deeper network architectures. The flexible framework developed in this work provides a solid foundation for such extensions.}    \label{fig:IndividualGBs1}
\end{figure*}

\begin{figure*}
    \centering
    \begin{subfigure}{0.45\textwidth}
        \centering
        \includegraphics[width=\textwidth, trim=0 0 0 0, clip]{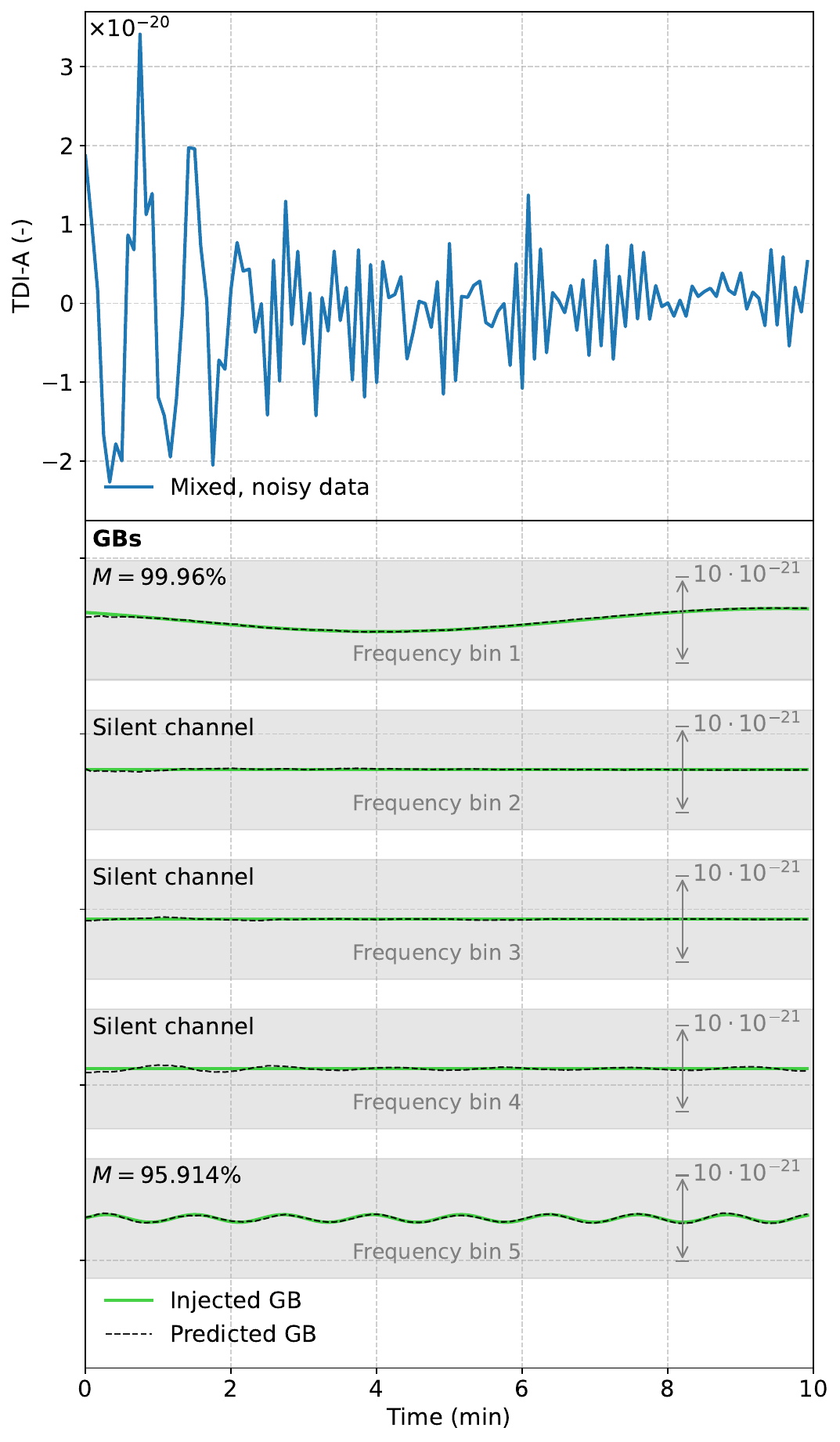}
        \caption{}
    \end{subfigure}
    \hfill
    \begin{subfigure}{0.45\textwidth}
        \centering
        \includegraphics[width=\textwidth, trim=0 0 0 0, clip]{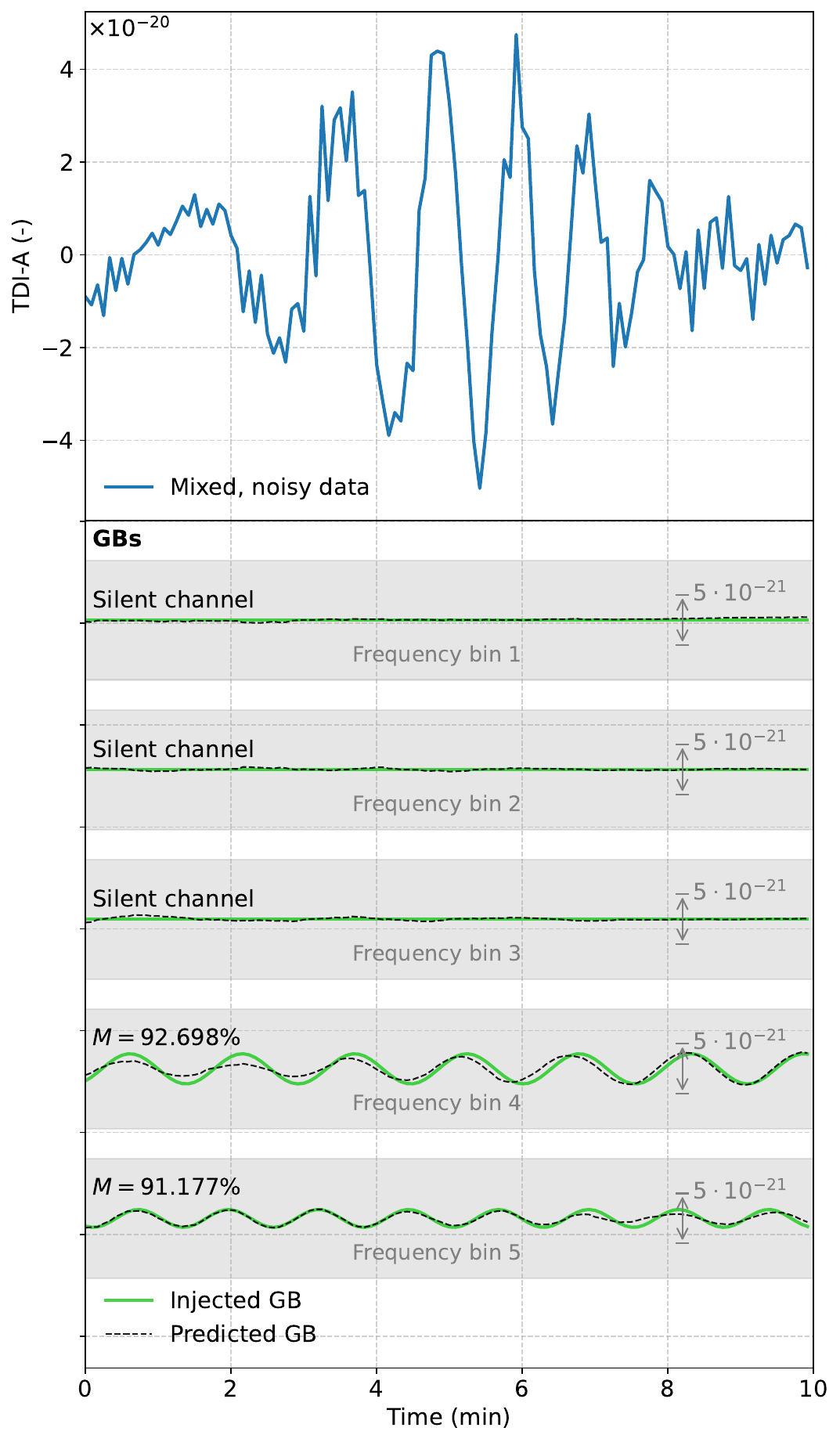}
        \caption{}
    \end{subfigure}
    \caption{\RaggedRight Comparison of the injected and predicted GB waveforms in the presence of a merging MBHB. We do not highlight the injected MBHB explicitly. }    \label{fig:IndividualGBs2}
\end{figure*}

Finally, we evaluate the performance of the GB decoder in distinguishing individual and overlapping GBs. Figures \ref{fig:IndividualGBs1} and \ref{fig:IndividualGBs2} compare the true and predicted number of GBs, demonstrating that the model accurately estimates the number of active sources even in the presence of glitches and merging MBHB. This supports the effectiveness of the frequency-bin approach. We notice that the outputs of silent channels are not entirely zero. This is expected, given the decoder design. In future iterations, we plan to refine the GB decoder head by incorporating a gating mechanism that learns to fully suppress inactive channels, minimizing spurious noise when no signal is present in a given bin.

\nh{To quantitatively assess the separation performance for GB signals, we perform an ensemble study analogous to Fig.~\ref{fig:match_vs_snr} using 150,000 synthetic LISA simulations. Each simulation includes a single variable-strength GB signal injected into a fixed realization of LISA-like noise, possibly including instrumental glitches and MBHBs. The GB signal is scaled using a range of amplitude factors, resulting in target signal-to-noise ratios (SNRs) between 1 and 100. For each mixture, the deep source separation model is applied to recover the GB waveform. Then, for each recovered GB waveform, we compute the match factor between the predicted and true signal as a function of the SNR. The results are given in Fig. Fig.~\ref{fig:Match_vs_SNR_GB}. The figure shows the distribution of match values across the ensemble, including the median (red line), interquartile range (dark shaded region), and the 5–95\% percentile range (light shaded region). The background density plot visualizes the individual scatter of simulations. At low SNRs, performance is limited, as expected, but the median match improves rapidly and exceeds 0.95 once the SNR reaches approximately 10.
In contrast to the analogous analysis performed with the MBHB decoder,  we observe a decline in the median match factor for GB signals at high SNRs. This decline reflects the model's behavior when confronted with out-of-distribution signals rather than an actual performance limitation. 

Note that a partial reason for the lower reconstruction accuracy of GBs compared to MBHBs and glitches lies in the relative amplitude differences between source types. Since the total loss is computed as a sum of equally weighted MSE terms, higher-amplitude sources like MBHBs tend to dominate the optimization. As a result, lower-amplitude GBs contribute less to the gradient signal and may be underfit. To address this imbalance, future work will explore adaptive loss weighting schemes, where a secondary network dynamically estimates task-specific weights based on source uncertainty or signal characteristics, allowing the model to balance reconstruction fidelity more uniformly across all components.

\begin{figure}[]
    \centering
    \hspace*{-1cm} 
    \includegraphics[clip, trim=0 0 0 0, width=0.5\textwidth]{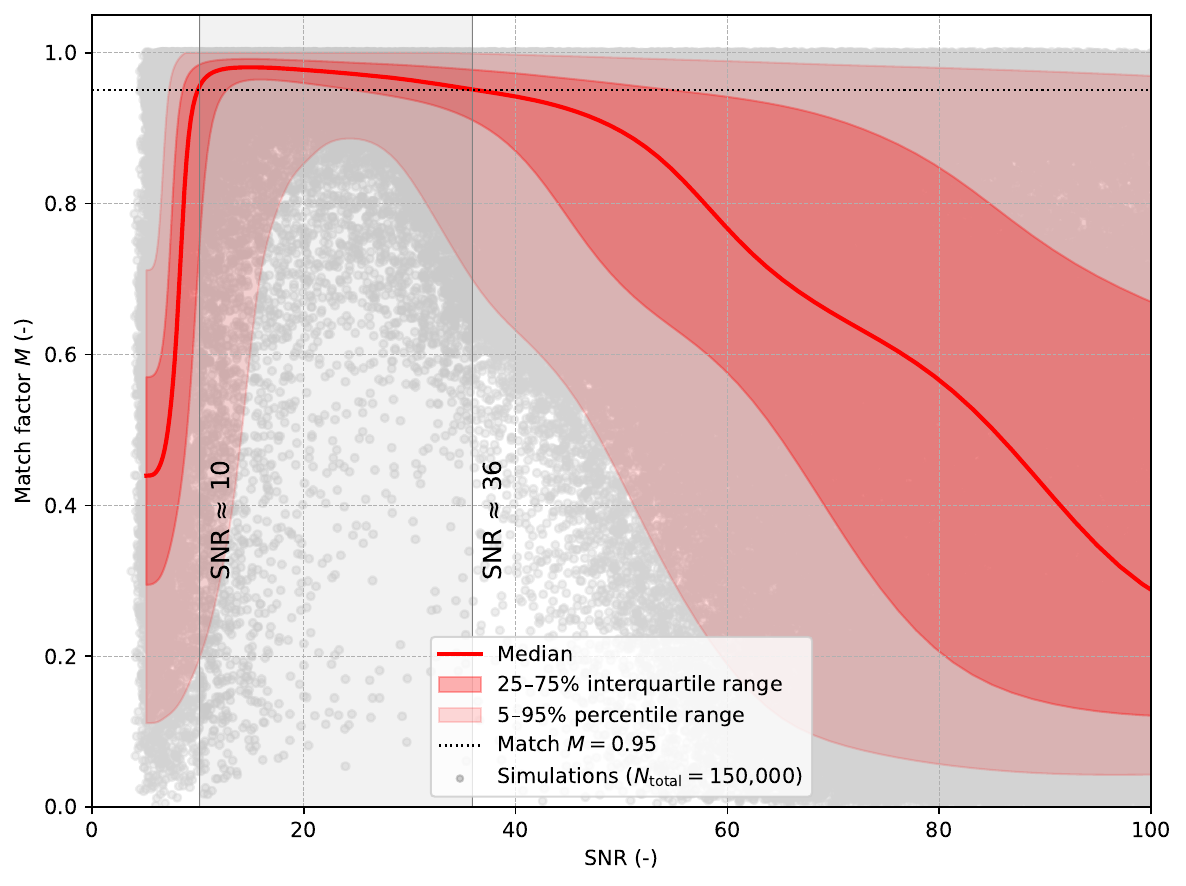}
    \caption{\RaggedRight
        Match factor $M$ as a function of SNR for isolated GB signals recovered from noisy TDI data. Gray points represent individual simulations, while red curves show the median and percentile ranges. The vertical line marks the SNR at which the median match first exceeds $M = 0.95$, a threshold chosen for visualization purposes. The drop in performance at high SNR reflects the model's response to out-of-distribution inputs.
    }
    \label{fig:Match_vs_SNR_GB}
\end{figure}

}

Scaling to the thousands of overlapping GBs anticipated in LISA's observations will require a more complex network architecture and larger training datasets. Additionally, integrating a multi-resolution transform, such as wavelets, into the encoder design may be essential. However, even with these enhancements, a decrease in GB separation performance is expected when analyzing realistic populations on small datasets. The expected decrease in resolution performance stems from the inherent challenges in resolving individual GB signals within a densely populated frequency spectrum. In the LISA frequency band, millions of GBs are expected to emit gravitational waves, leading to overlapping signals that create a confusion noise. This overlap makes it difficult to distinguish individual sources. Traditional methods for GB parameter estimation face similar issues, as they rely on resolving individual signals from a complex superposition of numerous sources. Consequently, longer observation times are necessary to improve the signal-to-noise ratio and to accurately infer waveform parameters. 

From this perspective, our model is designed with scalability in mind; we train on short data snippets and plan to apply a Demucs-like stitching procedure in future work to process arbitrarily long datasets once the model is trained. This approach involves segmenting long data streams into manageable, overlapping pieces, processing each segment individually, and then combining the results to reconstruct the full signal. This method has been effective in \texttt{demucs} and shows promise for application in gravitational wave data analysis. However, it's important to note that this stitching procedure is not part of the current study and will be explored in a follow-up investigation.

\nh{
\subsection{Non-stationarity and data gaps in long-duration inference}

In our current setup, the model is trained and evaluated on short data snippets of 2 hours in duration. Over such timescales, the Galactic binary population can be approximated as quasi-stationary since the individual sources evolve slowly and the overall structure of the foreground remains largely unchanged. Consequently, the training data -- and the latent representations learned by the encoder -- reflect only the stationary characteristics of the foreground within each segment. During inference, however, the model will eventually be applied segment-wise to arbitrarily long time series, and the outputs are combined using the aforementioned stitching procedure. This allows slowly varying foreground effects -- such as Doppler modulation and frequency evolution -- to emerge naturally in the reconstructed outputs.

We also note that the proposed framework is compatible with data gaps. Since each segment is processed independently, any valid data before and after a gap can be separately analyzed and reconstructed, with the gap manifesting as a discontinuity in the stitched output. If higher continuity is desired, established LISA gap mitigation strategies could be integrated at the preprocessing or postprocessing level, including time-domain gating, frequency-domain likelihood adaptation, or statistical inpainting methods. Alternatively, exposing the model to artificially gapped training data may enhance robustness in the presence of real interruptions. These directions offer promising extensions to improve the model’s applicability to realistic mission scenarios.}

\begin{figure}
    \centering
    \includegraphics[width=0.9\linewidth]{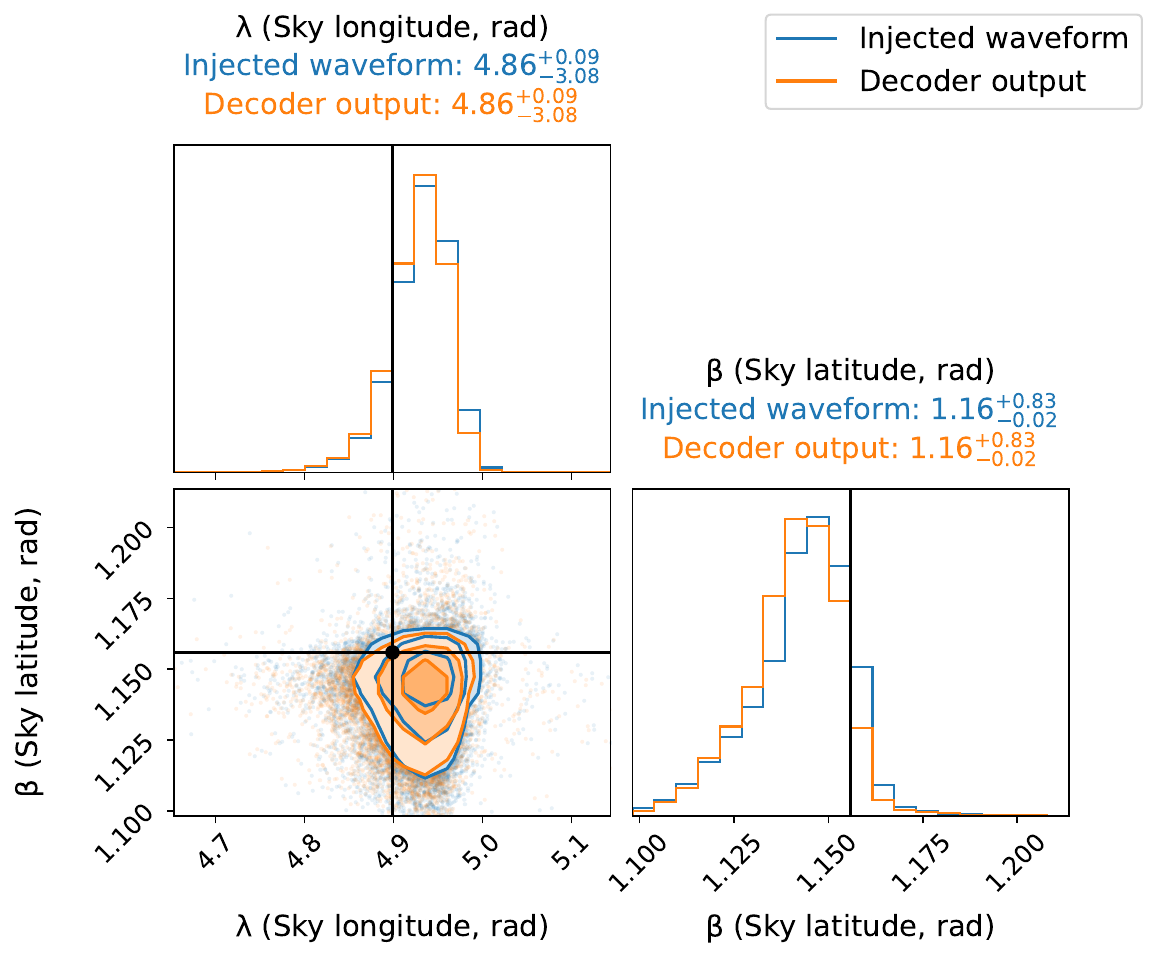}
    \caption{\RaggedRight 
    Posterior distributions from the SBI model trained on true injections, evaluated on both the true waveform and the high-match reconstructed waveform of Fig.~\ref{img:Example-SignalPrediction}. The two posteriors are nearly identical, indicating that accurate reconstruction preserves inference quality.}
    \label{fig:corner_87}
\end{figure}

\nh{

\subsection{Bayesian inference from separated signals}

A natural extension of the present framework involves estimating astrophysical source parameters from the separated signals. While the current model is designed for blind source separation, the resulting waveforms -- or alternatively, their latent representations -- could serve as inputs to Bayesian inference techniques aimed at recovering posterior distributions over source parameters. Simulation-based inference (SBI) offers a particularly promising approach in this context. By training neural density estimators on synthetic populations, one can learn a mapping from reconstructed signals to posterior distributions. When using the outputs of the decoder as inputs to the inference model, the SBI framework becomes implicitly aware of the separator's characteristics as it learns to account for any distortions introduced by the encoder-decoder architecture. This separator-aware formulation would ideally allow the posterior estimator to remain well-calibrated even when the signal reconstruction is imperfect. 

An alternative or complementary direction involves applying classical MCMC sampling directly to the raw TDI data. This remains the most principled approach when a well-defined likelihood function is available, especially since the noise characteristics are reliably known only for the original TDI data -- not for the decoded outputs. While such inference is computationally intensive, it could benefit from incorporating signal estimates from the separator as a preprocessing step: parameter point estimates obtained from the decoder outputs -- via matched filtering optimization or regression -- could be used to narrow the prior range, thereby accelerating burn-in and improving convergence behavior.

Looking ahead, the modular structure of the present framework naturally supports such extensions. In the present work, we take a first step in this direction by applying SBI to MBHB signals reconstructed from the deep source separation framework. We focus on two representative cases: the high-match MBHB example of Fig.~\ref{img:Example-SignalPrediction}, and the lower-match MBHB reconstruction of Fig.~\ref{fig:SignalPrediction_Overlap}(c). For now, the inference task is limited to sky localization.

We follow a two-step strategy. In the first step, we train an SBI model on true (injection-level) MBHB waveforms, enabling the neural density estimator to learn posteriors in an idealized, distortion-free setting. This model is then used to infer posteriors for both the true injected waveforms and the corresponding recovered waveforms from the separator. Comparing the two outputs reveals how separation accuracy affects inference: we expect that the high-match example yields a nearly identical posterior to the true case, while the lower-match example shows deviations due to waveform distortion.

In the second step, we train a new SBI model with equal architecture using the recovered waveforms themselves as input. This model becomes \textit{separator-aware}, learning to map distorted inputs to calibrated posteriors despite imperfections in signal reconstruction. We then compare its output -- obtained using the recovered waveform -- to the output of the original SBI model applied to the true injection. In the ideal case, both approaches yield comparable posteriors, demonstrating that separator-aware inference can recover the correct parameter distribution even in the presence of encoding and decoding artifacts.

A schematic overview of this pipeline is provided in Fig.~\ref{fig:sbi_flowchart}. While we focus here on MBHB sky localization, the methodology generalizes to other parameters and source classes.

\begin{figure}[t!]
\centering
\scalebox{0.9}{
\begin{tikzpicture}[node distance=1.3cm]

    \node (input) [startstop] {Noisy LISA data $x(t)$ };

    \node (dss) [process, below=of input, yshift=0.2cm] {
        Deep source separation \\ 
        \small $x(t) \rightarrow \big\{ \hat{x}_{\mathrm{MBHB}}(t),\, \hat{x}_{\mathrm{GB},i}(t),\, \hat{x}_{\mathrm{glitch}}(t) \big\}$
    };

    \node (sbi) [process, below=of dss, yshift=0.2cm, fill=orange!30]  {
        Simulation-based inference  \\
        \small $\hat{x}_{\mathrm{source}}(t) \rightarrow p(\theta_{\mathrm{source}} \,|\, \hat{x}_{\mathrm{source}})$
    };

    \node[rotate=90, anchor=center] at ([xshift=3.2cm]sbi.center) {\footnotesize One SBI per source type};

    \node (posterior) [startstop, below=of sbi, yshift=0.2cm] {
        Estimated posteriors $ p(\theta_{\mathrm{source}} \,|\, \hat{x}_{\mathrm{source}})$ 
    };

    \draw [arrow] (input) -- (dss);
    \draw [arrow] (dss) -- (sbi);
    \draw [arrow] (sbi) -- (posterior);
\end{tikzpicture}

}
\caption{\RaggedRight End-to-end separator-inference pipeline. Noisy LISA data is processed by the deep source separation model that disentangles the input into MBHB, Galactic binary, and glitch components. Each recovered signal is then passed to a source-specific simulation-based inference model to estimate physical parameters and infer posterior distributions.}
\label{fig:sbi_flowchart}
\end{figure}

The SBI model is built using the sequential neural posterior estimation framework provided by the \texttt{sbi} library \cite{BoeltsDeistler_sbi_2025}. Posterior distributions are modeled using neural spline flows, which offer a flexible and expressive class of density estimators. The prior is chosen as a uniform box distribution over the sky parameters $\lambda$ and $\beta$. Importantly, the training of the SBI model is fully decoupled from the training of the deep source separation network. We use 50,000 simulated examples to train the SBI model. This setup is intentionally kept simple and is not tuned for performance optimization; the goal is to facilitate controlled comparisons and assess the impact of reconstruction quality on downstream parameter estimation.

Figure~\ref{fig:corner_87} illustrates the result of the first-stage inference for the high-match case of Fig.~\ref{img:Example-SignalPrediction}. Here, we apply the SBI model trained on true injections to both the true MBHB waveform and the corresponding waveform recovered by the separator. The two posterior distributions are nearly identical in both sky latitude and longitude, demonstrating that when the reconstructed signal closely matches the true waveform, the downstream inference remains virtually unaffected. 

In contrast, Fig.~\ref{fig:corner_99} shows the same evaluation procedure applied to the lower-match example of Fig.~\ref{fig:SignalPrediction_Overlap}(c), where the MBHB waveform is partially distorted. In this case, the posteriors obtained from the true and recovered waveforms begin to diverge, particularly in sky longitude. This indicates that the inference quality degrades gracefully under moderate distortion and suggests room for improvement in modeling separator-induced uncertainty.

To mitigate this, we train the SBI model on the recovered waveforms instead of true injections. As shown in Fig.~\ref{fig:corner_99_corrected}, this separator-aware SBI framework successfully compensates for distortions introduced during source separation. The posterior obtained using the recovered waveform now aligns closely with the one from the true injection evaluated by the original SBI model. This demonstrates that the inference network can adapt to the reconstruction artifacts it is exposed to during training, effectively learning to 'undo' the distortions introduced by the encoder-decoder model. 

This initial integration of deep source separation with SBI serves as a proof of concept and outlook. Future work will extend this approach to a broader population of sources, incorporating full parameter estimation and systematic robustness studies across the LISA sensitivity range.
}

\begin{figure}
    \centering
    \includegraphics[width=0.9\linewidth]{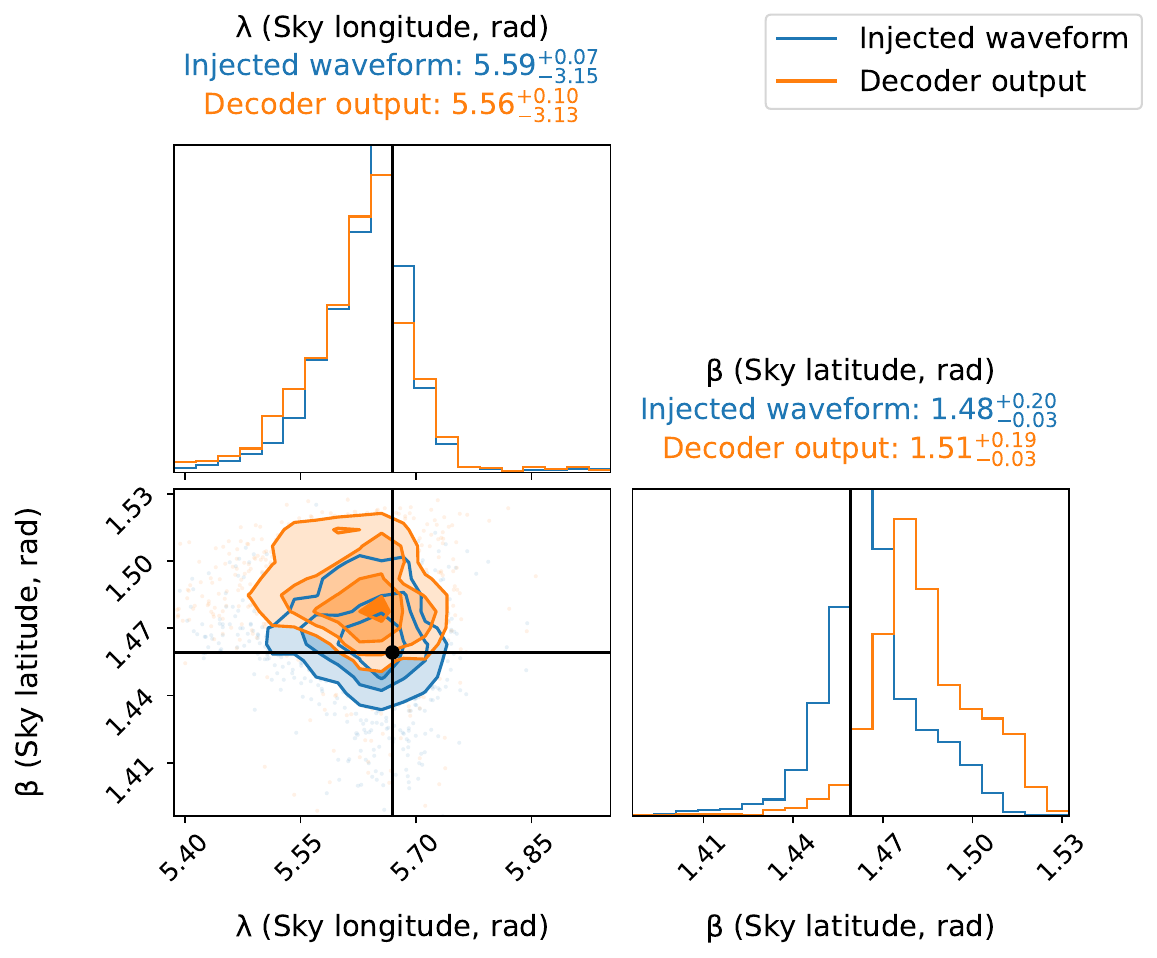}
    \caption{\RaggedRight 
    Posteriors obtained using the SBI model trained on true injections for the lower-match case of Fig.~\ref{fig:SignalPrediction_Overlap}(c). The distributions begin to diverge, particularly in sky longitude, reflecting the impact of signal distortion on downstream inference.}
    \label{fig:corner_99}
\end{figure}

\begin{figure}
    \centering
    \includegraphics[width=0.9\linewidth]{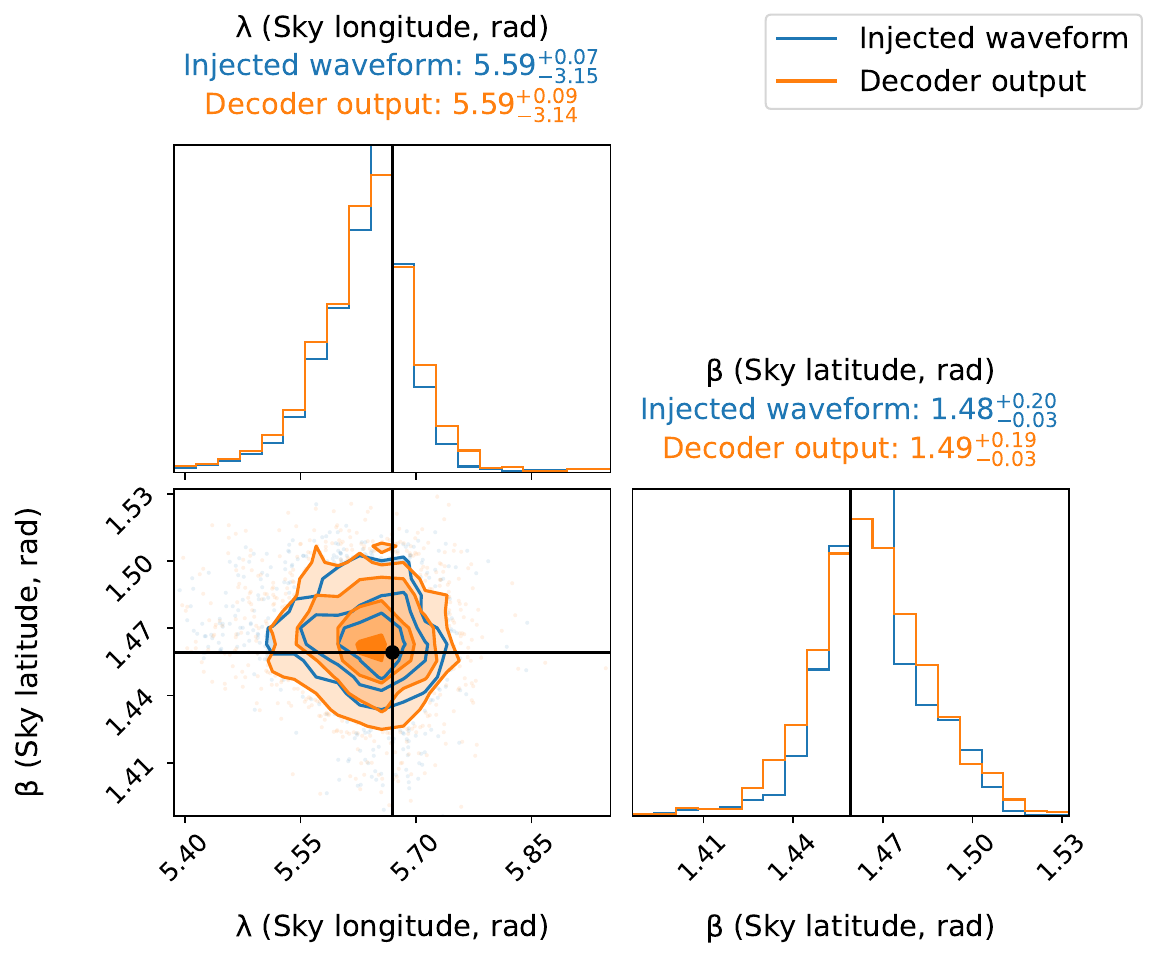}
    \caption{\RaggedRight 
    Posterior from the separator-aware SBI model trained directly on recovered waveforms, corresponding to the lower-match example shown in Fig.~\ref{fig:SignalPrediction_Overlap}(c). Despite distortions in the input, the inferred posterior matches the true-injection result from the standard SBI, demonstrating that the re-trained network has learned to compensate for reconstruction artifacts.}
    \label{fig:corner_99_corrected}
\end{figure}

\section{Conclusion}\label{sec5}

We presented a deep learning-based framework for blind source separation in high-dimensional LISA data, addressing overlapping gravitational-wave signals and non-stationary noise artifacts. Inspired by \texttt{demucs}, a model originally designed for audio processing, our approach employs a shared encoder-decoder architecture to disentangle complex signal components directly in a single step, bypassing iterative subtraction techniques, which represents a conceptual shift in methodology. The model isolates individual components dynamically through latent space clustering while remaining scalable to high-density astrophysical populations. \nh{A remaining limitation is that the current study restricts the degree of frequency overlap among individual GB sources, which will be addressed in future work.}

The evaluation of our method on simulated LISA data demonstrated its ability to successfully handle challenging observational conditions, including high-redshift MBHB mergers embedded in realistic noise, overlapping glitches, and quiet periods devoid of mergers, thereby minimizing false positives.

Our current implementation is a proof-of-concept study, that restricts the number of overlapping GBs and excludes complicated EMRI waveforms. Nevertheless, the results demonstrate the potential of deep source separation for LISA data analysis. Notably, even with this simple framework -- consisting of a few hundred lines of code and trained on a modest dataset -- model inference operates efficiently on a standard laptop within seconds. \nh{Further training and implementation details are provided in Appendix \ref{app:training}.} Building on the presented results, we will explore the framework's scalability to more complex, large-scale source populations. We are optimistic about its broader applicability.

In parallel, we have taken initial steps to integrate SBI into the analysis pipeline. Leveraging neural density estimators trained on synthetic waveforms, SBI enables direct posterior estimation from the separated signals, even in the presence of reconstruction artifacts. This approach is particularly valuable for enabling fast, calibrated parameter estimation without requiring a full likelihood model. As shown in our proof-of-concept results, separator-aware inference models can learn to compensate for distortions introduced by the encoder-decoder architecture, maintaining robust performance across a range of reconstruction qualities. Looking ahead, we envision this integration as a foundation for end-to-end gravitational-wave analysis pipelines, where source separation and parameter inference are tightly coupled within a unified learning-based framework.

As LISA's launch approaches, scalable and efficient data analysis methods become increasingly important. Deep source separation offers a promising avenue for addressing the mission's low-latency and global fit requirements, complementing traditional Bayesian inference and MCMC techniques. By refining and extending the method presented in this work, we aim to drive the development of next-generation gravitational wave detection strategies, setting a new standard for ML-based data analysis in our astrophysics community.

\nh{
\appendix
\section{Training setup and model complexity}\label{app:training}
The total number of trainable parameters in the Demucs-style multi-source separation model is approximately 2.79 million. This figure reflects the combined weights of the shared encoder, bottleneck, and three decoders targeting MBHBs, glitches, and Galactic binaries (GBs), respectively. The GB decoder outputs one signal per predefined frequency bin; in this work, we use five such bins.

The deep source separation model was trained on a dataset of 25{,}000 simulated time-domain mixtures, each comprising MBHB signals, glitches, and multiple GB sources embedded in instrumental noise. Each training segment corresponds to a 2-hour duration.

Training was conducted on a MacBook Pro with an M2 Max chip and 32\,GB of unified memory, using PyTorch. The model was trained for 25 epochs, requiring approximately 4 hours.

This configuration achieved satisfactory source separation performance for initial evaluations. Larger-scale experiments using GPU-accelerated hardware are underway to assess scalability with increased training data and more expressive model architectures.
}

\section*{Acknowledgments}
This research was funded by the Gravitational Physics Professorship at ETH Zurich. The author thanks Michele Vallisneri for the insightful discussions and for his valuable contributions to editing the manuscript. Gratitude is also extended to the LISA Simulation Working Group and the LISA Simulation Expert Group for their engaging exchanges on all simulation-related activities. Copyright 2025. All rights reserved.


\bibliography{apssamp}

\end{document}